\documentclass[prd]{revtex4-1}
\usepackage{bm}
\usepackage{latexsym}
\usepackage{amsfonts,amssymb}
\usepackage{amssymb}
\usepackage{graphicx,epsfig}
\usepackage{amsmath,amssymb}
\def\be{\begin{equation}}
\def\ee{\end{equation}}

\newcommand{\ud}{\textrm{d}}
\newcommand{\vev}[1]{\left<\, #1\,\right>}

\begin{document}

\title{On 1-loop diagrams in AdS space and the random disorder problem}
\author{Ling-Yan Hung, Yanwen Shang}
\email{jhung,$\,$yshang@perimeterinstitute.ca}
\affiliation{Perimeter Institute for Theoretical Physics, Waterloo, Ontario N2L
2Y5, Canada}

\begin{abstract}
We study the complex scalar loop corrections to the
boundary-boundary gauge two point function in pure AdS
space in Poincare coordinates, in the presence of a boundary quadratic
perturbation to the scalar. These perturbations correspond to
double trace perturbations in the dual CFT and modify the boundary
conditions of the bulk scalars in AdS.  We find that,
in addition to the usual UV divergences, the 1-loop calculation suffers from
a divergence originating in the limit as the loop vertices approach
the AdS horizon. We show that this type of divergence
is independent of the boundary coupling, and making use of which
we extract the finite relative variation of the imaginary
part of the loop via Cutkosky rules as the boundary perturbation
varies. Applying our methods to compute the
effects of a time-dependent impurity
to the conductivities using the replica trick in AdS/CFT,
we find that generally an IR-relevant disorder reduces the
conductivity and that in the extreme low frequency limit the
correction due to the impurities overwhelms the planar CFT result even
though it is supposedly $1/N^2$ suppressed.  Comments
on the effect of time-independent impurity in such a system are
presented.
\end{abstract}

\maketitle


\section{Introduction}

AdS/CFT correspondence\cite{Maldacena:1997re,Gubser:1998bc,Witten:1998qj}
 has proven itself an extremely powerful tool in
extending our understanding of strongly coupled quantum field theories,
stretching its influence into many different realms of physics.
Particularly, there has been a surge of interest in applying
these techniques
in condensed matter systems in recent days
(see for example a recent review \cite{Hartnoll:2009sz}
and references therein).
Thus far, studies of AdS/CFT have predominantly focused on extracting
the leading large $N$ physics in the CFT via (semi)-classical
supergravity computations
in AdS. However, if one attempts to make contact with more
realistic systems, $N$ should be finite and there exist many
circumstances in which $1/N$ correction is important.
These corrections correspond to
non-planar diagrams in the CFT, and quantum loop corrections
in the AdS bulk, and in particular 1-loop diagrams in AdS give
rise to $1/N^2$ suppressed corrections.
A number of physical phenomena are known to show up only if
one includes loop
corrections in the bulk, such as the de Haas - van Alphen
quantum oscillations \cite{Denef:2009kn,Denef:2009yy} in
strongly coupled charged systems, hydrodynamic long-time tails of
a fluid \cite{CaronHuot:2009iq}, and the holographic
manifestation of the Mermi-Wagner therem \cite{Anninos:2010sq}.
More recently it is also shown in \cite{Mross:2010rd}
circumstances where the loop corrections can compete with the
tree contribution in certain non-fermi-liquids.
Consequently, there has been growing interest in the
community in understanding quantum loops in the supergravity theory in AdS.
Calculating loops in AdS space could entail extra
complexities, such as additional divergences, due to the
non-trivial properties of the geometry. It certainly
is of importance to investigate
this problem closely, clarifying possible
 obstacles and extracting
physical implications.

As one of the original motivations of the current study,
it has been proposed in \cite{Fujita:2008rs}
that one may incorporate the replica trick that is commonly adopted in the
context of condensed matter systems to capture the effect of random disorder,
through AdS/CFT correspondence.
There, one introduces replicas of the AdS spaces and the bulk fields.
The coupling to the random disorder appears on the AdS side
as boundary terms that
couple to the replicated bulk fields, thus effectively changing the boundary
conditions those fields satisfy.  These kind of boundary perturbations
have been studied previously for example in \cite{Witten:2001ua,
Aharony:2001pa,Berkooz:2002ug,Aharony:2005sh,
Papadimitriou:2007sj,Hartman:2006dy,Elitzur:2005kz},
and they are shown to correspond to multiple trace, and in case of a
quadratic boundary term double trace perturbations, in the dual CFT theory.
It is also shown there that the effect of these boundary
perturbations of a scalar field for example, are only
mediated to other fields, such as the graviton or the photon,
beginning at 1-loop order.
This implies that the effect of disorder on
transport coefficients such as conductivities only shows up in
loops \footnote{One might wonder why the effect of disorder is so weak.
This is actually related to the fact that in the replica trick implemented
in the bulk we have implicitly assumed that the strength of the
coupling to the random disorder is only of order $1/N^2$.
See for example the discussion in \cite{Aharony:2001pa}. }.

Motivated by these interesting studies, we take a phenomenological
approach and consider charged scalar fields
coupled to $U(1)$ gauge fields in pure AdS space in a general
Einstein-Maxwell theory in $d+1$ dimensions, and study particularly the
scalar 1-loop Witten diagrams correcting the photon boundary-boundary 2-point
function. We then read-off the conductivity of
the dual theory via the usual AdS/CFT dictionary.
These scalars are allowed to satisfy mixed boundary
conditions, corresponding to the introduction of double trace boundary
perturbations as mentioned above. It is found that the
computation done in Poincare coordinates in Lorentzian signature suffers
from a divergence that arises as both vertices in the loop approach the
horizon (or the IR limit) where the geodesic distance between
them vanishes. It is surprising, however, that such a divergence persists
even within the imaginary part of the loop.  More specifically, it
manifests itself as a singularity in the integral over loop-momenta in
the collinear limit\footnote{This limit, however is quite different from
the usual infrared divergence in the loop correction
of the photon propagator, which appears as the external
invariant $d$-momentum goes on-shell i.e.$p_\mu p^\mu\to0$ .
The singularity here occurs even for finite external
invariant $d$-momentum. }, that
cannot be cured even though the phase-space volume approaches zero there.
These divergences, however, turn out to be independent of the
strength of the boundary perturbation.  We demonstrate this property
by studying the asymptotics of the propagators in the near horizon
limit and extract the precise forms of these singularities, even though
the full analytic result of the loop integral is not generally expressible
in terms of elementary functions whose properties are otherwise obscure.
We obtain in this way both the coefficients and the powers of
each singular terms analytically and find that they have simple
dependence on the number of bulk dimensions.

Given the universal nature of these divergences, we
extract finite results by considering
the differences of diagrams evaluated at different
 boundary couplings.
We managed to
compute the relative conductivities for a wide range of
boundary couplings
$f$ and find that they interpolate smoothly between the two conformal
limits $f\to 0$ and $f\to \infty$. We then return to the study of
random disorder via the replica trick, and compute the
conductivities under the influence
of random impurities.  At $d=2+1$, we give an example where
the computation can be done exactly. We find that the presence
of the impurities generally reduce the conductivity.
Moreover in the low frequency limit the correction overwhelms
the planar CFT result. This implies that a re-summation in the
deep IR is probably necessary, but
also suggests a possible resolution to the puzzle of how
$1/N$ suppressed corrections
could actually drastically change the IR behavior of the
transport coefficients. A more complete
analysis of these re-summation is however not pursued
in the present paper.

The organization of the paper is as follows. In section \ref{sec:basics}
we present the
form of scalar bulk-to-bulk propagators satisfying general mixed
boundary condition at the AdS boundary. In section \ref{sec:vertex}
we briefly review the vector bulk-to-boundary
propagator in momentum space and the scalar-vector vertices that are
relevant for the loop diagrams. We also introduce Cutkosky
rules in AdS space and compute the imaginary part of the scalar
1-loop correction to the photon
boundary-boundary 2-point function.  We did not include
similar contribution of other fields such as gravitons because we are
mainly interested in the leading dependence on the coupling of
the boundary perturbation. We will show that a divergence arises
but is independent of the boundary perturbation, and that
the remaining finite part interpolates smoothly between the two
conformal limits. We also make some comments on these divergences more
generally in other loops in AdS space.
In section \ref{sec:disorder} we apply our method in the context of
random disorder. We conclude our results in section \ref{sec:conclusion}.
Further details of the computations are relegated to the appendices.

\section{Scalar propagators of mixed boundary conditions}
\label{sec:basics}
Consider AdS space in $d+1$ dimensions with the metric
\be
ds^2 = \frac{1}{z^2} (- dt^2 + dz^2 + \sum_{i}^{d-1}dx_i^2),
\ee
the Green's function of scalar fields in AdS space of
mass $m$ satisfy the inhomogenous Klein-Gordon
equation sourced by delta-functions:
\be
\label{eq:greens}
\left(\frac{1}{\sqrt{g}}(\partial_{z_1} g^{zz}\sqrt{g}
 \partial_{z_1})  + g^{\mu\nu}\partial_\mu\partial_\nu
 - m^2 \right) G_{\Delta_{-}}(z_1,z_2,x_1, y_1) =
 \frac{1}{\sqrt{g}}\delta^d(x_1-y_1)\delta(z_1-z_2).
\ee
In general these scalar fields behave in the boundary limit
$z \to 0$ as
\be \label{bexp}
\phi(z) \sim \alpha z^{\Delta_+} + \beta z^{\Delta_-},\qquad
\Delta_\pm = \frac{d}{2} \pm \nu,
\ee
where we have defined
\begin{equation}
\nu=\sqrt{m^2 + \frac{d^2}{4}}\,.
\end{equation}
When both $\Delta_{\pm}$ are greater than zero and thus the
corresponding wave-function normalizable, one can
insert boundary terms of the form
\be\label{bterms}
\delta S_{\partial \mathcal{M}}= \int_{\partial \mathcal{M}}
f \beta^2,
\ee
so that the variation of the action only vanishes on the boundary
if the scalar field satisfies the following boundary condition
\be \alpha = f \beta. \label{bc1}
\ee
For finite $f$ the dual operator in the CFT whose vev is given
by $\beta$ has conformal dimension $\Delta_-$ in the UV, and
therefore the dimension of $f$ is given by
\begin{equation}
[\,f\,]=d-2\Delta_-=2\nu\,.
\end{equation}
The unitarity condition and the requirement of preserving
conformal symmetry in the UV, i.e. the perturbation is irrelevant
in the UV, demand $|\nu|<d/2$, or $-d^2/4\le m^2 < 0$ \cite{Klebanov:1999tb}.

These boundary terms are multi-trace perturbations
from the point of view of the dual CFT, and have been first studied in e.g.
\cite{Witten:2001ua,Aharony:2001pa,Berkooz:2002ug,
Aharony:2005sh,Papadimitriou:2007sj,Hartman:2006dy,Elitzur:2005kz}.

The scalar Klein-Gordon equation can be readily solved
in momentum space of the flat directions. For given
$d$-momentum $k^\mu$, $x^{d/2} J_\nu (k z)$ is a solution
to the homogeneous version of equation \eqref{eq:greens}
i.e. without the $\delta$-function source,
where $J_\nu(k z)$ is a Bessel function of order $\nu$ and $k=|k^\mu|$.
One can easily solve for the Green's function making use of the completeness
of Bessel functions for given boundary condition at $z\rightarrow 0$.
Instead of piecing two solutions defined for $z_1>z_2$ and $z_1<z_2$ as
in \cite{Gubser:2002zh}, we can  express $G_{\Delta_-}(z_1,z_2,k)$
as the following integral~\footnote{Another
straightforward way of obtaining this propagator
is expanding all the fields in terms of the properly chosen radial
wave-functions and express the action in terms of the expansion coefficients.
The inverse of the resultant kinetic term in the action leads
immediately to equation \eqref{b2b}.}
\begin{equation}\label{b2b}
G_{\Delta_-}(z_1,z_2,k) = \int_0^{\infty} d\Lambda \frac{\Lambda}{\Lambda^2 + k^2}
J_{\nu, f}(\Lambda,z_1)J_{\nu, f}(\Lambda,z_2),
\end{equation}
where
\begin{equation}
\label{besself}
J_{\nu, f}(\Lambda, z)\equiv N(\Lambda)
	[A(\Lambda)J_\nu(\Lambda z)+B(\Lambda) J_{-\nu}(\Lambda z)]
\end{equation}
is the linear combination of both Bessel functions of order $\nu$,
properly chosen to satisfy the boundary conditions \eqref{bc1}
at $z\rightarrow 0$. It is straightforward to check that we must choose
\be
A = 1, \qquad B = \frac{(2\Lambda)^{2\nu}}{f}
\frac{\Gamma[1-\nu]}{\Gamma[1+\nu]},
\ee
and the corresponding normalization factor
\begin{equation}
N(\Lambda)^2 = \frac{1}{1 + 2B(\Lambda)\cos(\nu\pi)+ B(\Lambda)^2}\,,
\end{equation}
such that $J_{\nu, f}$ is normalized in the sense that
\begin{equation}
\int_0^\infty z\Lambda_1 J_{\nu, f}(\Lambda_1, z)
J_{\nu, f}(\Lambda_2, z)\ud z
=\delta(\Lambda_1-\Lambda_2)\,.
\end{equation}
The expression for $N(\Lambda)$ can be justified most easily by checking
the asymptotic forms of $J_{\nu}$
and $J_{-\nu}$, or using the orthogonality properties
of the standard Bessel functions.

One could check,  as we demonstrate in Appendix \ref{app:contour},
that this representation agrees with the bulk-to-bulk propagator
in \cite{Gubser:2002zh} obtained in the Euclidean signature, which
simply means $k^2$ is positive in (\ref{b2b}).
The virtue of making use of this representation
will be made manifest when we begin computing loop corrections, where we are
spared of the difficulty of dealing with the step function $\Theta(z-z')$ present in the representation in \cite{Gubser:2002zh}.
Since we are interested in Lorentzian signature we will
have to specify precisely the causal
structure of our propagators. In the following analysis,
we will consider loop corrections to a retarded
correlation function. Feynman, or time-ordered,
propagators, as discussed in \cite{Chalmers:1998wu},
can be obtained by a simple $i \varepsilon$ prescription.
Namely, one makes the replacement
\be
\frac{1}{\Lambda^2 + k^2} \to \frac{1}{\Lambda^2 + k^2 - i \varepsilon}.
\ee
Similarly when we consider retarded (advanced) propagators,
we will put all the poles of $k_0$ in the lower (upper) half complex
$k_0$ plane.

\section{1-loop correction to gauge two-point correlation and
Cutkosky rules in AdS space}
\label{sec:vertex}

We are interested in this paper the charged scalar 1-loop correction to the
boundary-boundary correlator of a $U(1)$ gauge field in
$\textrm{AdS}_{d+1}$.  According to the standard AdS/CFT correspondence,
a $U(1)$ conserved current in a $d$-dimensional CFT is generally
dual to a $U(1)$ gauge field, or the bulk ``photon'' in
a $d+1$ dimensional AdS space. One can view that the CFT lives
on the AdS boundary and there is the relation
\begin{equation}
\vev{j_{\mu} j_\nu}_{\textrm{CFT}_d}=
\vev{A_\mu A_\nu}_{\partial\textrm{AdS}_{d+1}}\,.
\end{equation}

We will be considering, on the AdS side, loop corrections to this
correlator from a charged minimally coupled scalar field that satisfies
the general mixed boundary condition explained above. On the CFT side,
these loop corrections correspond to $1/N^2$ correction to
$\vev{j_\mu j_\nu}$.

\subsection{Interaction vertices and the photon propagator}
Let us first briefly review the photon boundary-to-bulk propagator at
tree level and its interaction with the complex scalar.
For convenience, we will work in the gauge
\be
A_z =0 \qquad \partial_\mu A^\mu =0.
\ee
In Euclidean signature, the boundary-to-bulk propagator
satisfying the usual Dirichlet boundary condition at $z=0$ in this gauge is
\cite{D'Hoker:1999jc, Liu:1998bu,Liu:1998ty,Chalmers:1998wu}
\be\label{photon_prop}
A_\mu(z,p) = J^{\perp}_\mu(p)
\frac{(pz)^{d/2-1}K_{(d/2-1)}(p z)}{(p\epsilon)^{d/2-1}K_{(d/2-1)}(p \epsilon)},
\qquad p^\mu J^{\perp}_\mu(p)=0.
\ee
The propagator has been normalised and $\epsilon$ is the UV
(or boundary $z\to 0$) cut-off.
The boundary source $J^{\perp}$ satisfies the transverse condition
as a result of the gauge choice. The Lorentzian propagator can be
readily obtained from the Euclidean one by analytic continuation.
Depending on whether one is interested in {\it in-coming} or {\it out-going}
 boundary conditions at the horizon $z\to \infty$, one could replace
 $p$ by $\pm i p$ accordingly. Since we are computing a retarded
 correlation function, we will take one external leg to satisfy
 in-going boundary conditions, whereas the other leg should be
 the complex conjugate
\footnote{For a detailed discussion see for example
\cite{Son:2002sd, Herzog:2002pc,Skenderis:2008dg}}.

There are two types of interaction vertices between
minimally coupled charged scalars and photons.
The 3-point and the 4-point vertices are given by
\begin{eqnarray}
\label{eq:vertices}
V_{A\phi^\dagger\phi}&=& \int d^dx dz \sqrt{-g}
 g^{\mu\nu} (-i)A_\mu (\phi^\dagger
 \partial_\nu\phi-\phi \partial_\nu\phi^\dagger ) \nonumber \\
V_{AA\phi^\dagger\phi}&=& \int d^dx dz \sqrt{-g}
g^{\mu\nu} A_\mu A_\nu \phi^\dagger \phi.
\end{eqnarray}

Given that the sources $J^\perp$ in our gauge have to
satisfy the transverse
condition, only the component satisfying the Ward-identities
$p^\mu\vev{j_\mu j_\nu}=0$ would contribute. For general external
momenta $p$ the 1-loop contribution is not expected to satisfy the
$d$-dimensional flat-space Ward identities for the boundary
theory. When $p$ approaches zero however, the photon wave function
becomes independent of the radial coordinate and the integral over
the radial position of the vertices yield simple delta functions.
In this limit the amplitude is reduced to a $d$-dimensional calculation,
and can be shown to be transverse after
all 1-loop diagrams are combined as usual in flat-space.

\subsection{1-loop diagrams and the Cutkosky rules in AdS space}
\begin{figure}[h!]
\includegraphics[width=0.6 \textwidth]{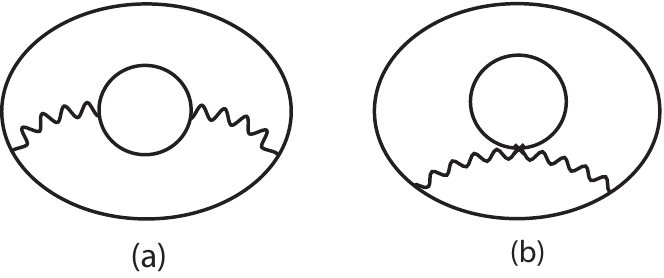}
\caption{\label{wittendiagram}}
\end{figure}

Now we turn to the 1-loop correction to the boundary-boundary
correlator of the bulk photon.  Our main objective is
to find their imaginary contributions.
There are two loop diagrams relevant to the current study as shown in figure
(\ref{wittendiagram}).
Both diagrams contribute to the $1/N^2$ correction to
the current-current correlation on the CFT side. Let
us denote the contribution from the diagram~(a)
as $\vev{j_\mu j_\nu}'$. Using the propagator \eqref{b2b} and
\eqref{bulk-boundary}, this diagram boils down to the following integral
\begin{equation}
\label{eq:loop1}
\vev{j_\mu(-p) j_\nu(p)}'
	=\frac{-P_{\mu\alpha}(p)P_{\nu\beta}(p)}
	{\epsilon^{d-2} |K_{d/2-1}(ip\epsilon)|^2}\int
\frac{\ud^d k\ud\Lambda_1\ud\Lambda_2}{(2\pi)^d}
	\frac{(p-2k)_\alpha(p-2k)_\beta \; \Lambda_1\Lambda_2
		|I_{\nu, f}(i p, \Lambda_1, \Lambda_2)|^2}
	{(\Lambda_1^2+k^2)_A(\Lambda_2^2+(p-k)^2)_R}\,,
\end{equation}
where we have included the only non-vanishing
contribution to this loop, involving the
product of a retarded and an advanced scalar
propagator. Our readers are reminded of the
respective pole structure of the propagators
by the subscripts $R$ and $A$ above.
We imposed the gauge condition by introducing the usual
projection operator
\begin{equation}
P_{\mu\nu}(p)=\eta_{\mu\nu}-\frac{p_\mu p_\nu}{p^2}\,.
\end{equation}
We have also defined the ``vertex function''
$I_{\nu, f}$ as
\begin{equation}
\label{eq:def_inuf}
I_{\nu, f}(p, \Lambda_1, \Lambda_2)\equiv
\int_0^\infty \ud z \, z^{\frac{d}{2}}K_{\frac{d}{2}-1}(p z)
	J_{\nu, f}(\Lambda_1, z) J_{\nu, f}(\Lambda_2, z)\,,
\end{equation}
which can be read-off directly from equation \eqref{eq:vertices}
expressed in momentum space for the $d$ flat directions.

More generally, it is possible that the two scalars
propagating in
the loop satisfy different mixed boundary conditions,
in which case
one would have to generalize the ``vertex function''
in the loop-integral as
\begin{equation}
\label{eq:def_inuf1f2}
I_{\nu, f_1, f_2}(p, \Lambda_1, \Lambda_2)\equiv
\int_0^\infty \ud z \, z^{\frac{d}{2}}K_{\frac{d}{2}-1}(p z)
	J_{\nu, f_1}(\Lambda_1, z) J_{\nu, f_2}(\Lambda_2, z)\,.
\end{equation}
This function has an analytic expression that will be discussed further
below. At the moment, we just note that,
following immediately from the properties
of the Bessel functions,
\begin{equation}
I_{\nu, f_1, f_2}(e^{i\frac{\pi}{2}} p, \Lambda_1, \Lambda_2)
	=I^\dag_{\nu, f_1, f_2}(e^{-i\frac{\pi}{2}}p, \Lambda_1, \Lambda_2)\,,
\end{equation}
which we have already made use of in equation \eqref{eq:loop1}.

Clearly, the integral would suffer from divergence near the boundary
$z\to 0$, interpreted as a UV divergence, if $2|\nu|>1$. To avoid
complications, we will restrict our attention in this paper to
$|\nu|< \frac{1}{2}$.

Written in this form, integral \eqref{eq:loop1}
can be conveniently interpreted in the $d$-dimensional language
as the total contribution to the 1-loop correction of
$\vev{A_\mu A_\nu}$
from all pairs of KK-scalars
of masses $\Lambda_1$ and $\Lambda_2$. These KK-scalars form a
continuous
infinite tower whose mass take all values from $0$ to infinity,
as should be expected since the $z$-dimension is non-compact.
Their contribution to the loop is integrated over,
convoluted by a non-trivial
``vertex function'', given by $I_{\nu, f}(i p, \Lambda_1, \Lambda_2)$,
or more generally $I_{\nu, f_1, f_2}$,
describing the mixing of the KK-scalars and their coupling to the gauge field.
It should be mentioned that the vertex function
$I_{\nu, f_1, f_2}$ as defined in
\eqref{eq:def_inuf} is not a proper integral and needs to be regularized.
Most naturally, it is regularized by adding to the external momentum $p$
a small positive imaginary part $i \varepsilon$ so
that \eqref{eq:def_inuf} becomes well-defined since its integrand
decays exponentially as $z\rightarrow +\infty$, and then taking the
$\varepsilon\rightarrow 0$ limit. The result is in general
finite and well-defined, but may contain various singularities
depending on the values of $(p, \Lambda_1, \Lambda_2)$. Some of
the singularities are severe enough that they can render the loop
integral \eqref{eq:loop1} ill-defined. Below, we will devote a
significant part of this paper discussing these singularities
and the regularization of the loop integral.

In this $d$-dimensional language, one can readily obtain
the imaginary part of the loop \eqref{eq:loop1} using the
Cutkosky Rules, or the analog of the``optical theorem'' for S-matrices.
At the level of tree-diagrams this has
been considered in \cite{Chalmers:1998wu}. Analyzing the pole
structure for the product of retarded and advanced propagators
in the same vain shows that the imaginary part can be evaluated
by putting the propagators in the loop on-shell, and, as in
Feynman loops, only the physical poles would contribute.
The cut diagram corresponds
to the amplitude of a single ``photon'' decaying into two
 KK-scalars of masses $\Lambda_1$ and $\Lambda_2$
respectively.  One then finds the modulus of the amplitude and integrates
it over the phase space allowed by the kinematics as well as all
positive values of $\Lambda_1$ and $\Lambda_2$.
Equivalently, one can replace the scalar propagators
$[(\Lambda_1+k^2)(\Lambda_2+(p-k)^2)]^{-1}$ by their on-shell
conditions, i.e.
$-(2\pi)^2\delta(\Lambda_1^2+k^2)\delta(\Lambda_2^2+(p-k)^2)$,
and carry
out the remaining integral.  The presence of the two $\delta$-functions
confines the integral into a compact region in the phase space. In fact
the result is only non-vanishing if the incoming momentum
$p$ is time-like and $\sqrt{p_0^2-|\mathbf{p}|^2}\ge \Lambda_1+\Lambda_2$.

Consider the simple example where $p_\mu = (p_0,0, ..., 0)$ and evaluate
the spatial components of the loop diagram. By rotational symmetry
and via the Cutkosky rules, we find
\begin{eqnarray}
&&\int \frac{d^dk}{(2\pi)^{d-2}} k_i k_j\delta(\Lambda_2^2
+ (p-k)^2)\delta(\Lambda_1^2 + (k)^2)  \nonumber \\ &&=\delta_{ij}
\frac{\Omega_{d-2}}{2(d-1)(2p_0)^{d}(2\pi)^d}
\left((p_0^2 - (\Lambda_1 + \Lambda_2)^2)(p_0^2 -(\Lambda_1 - \Lambda_2)^2)
\right)^{(d-1)/2} \nonumber \\
&&=\delta_{ij}\frac{\Omega_{d-2} p_0^{d-2}}{8(d-1)(4\pi)^{d-2}}
\left((1 - (\tilde{\Lambda}_1 + \tilde{\Lambda}_2)^2)(1 -(\tilde{\Lambda}_1
- \tilde{\Lambda}_2)^2)\right)^{(d-1)/2},
\end{eqnarray}
where $\Omega_{m}$ is the volume of an $m$-sphere and
$\tilde\Lambda_{1,2}\equiv\Lambda_{1,2}/p$.
The time-time component of the momentum integral similarly evaluates to
\begin{eqnarray}
&&\int\frac{d^dk}{(2\pi)^{d-2}} (p_0-2k_0)^2
\delta(\Lambda_2^2 + (p-k)^2)\delta(\Lambda_1^2 + (k)^2) \nonumber \\
&&=\frac{\Omega_{d-2}p_0^{d-2}}{8(4\pi)^{d-2}}(\tilde{\Lambda}_1^2
		-\tilde{\Lambda}_2^2)^2\left((1 - (\tilde{\Lambda}_1
		+ \tilde{\Lambda}_2)^2)(1 -(\tilde{\Lambda}_1 -
	\tilde{\Lambda}_2)^2)\right)^{(d-3)/2}\,.
\end{eqnarray}
Putting the pieces together, we are left with the integrals
of $\tilde \Lambda_1$ and $\tilde \Lambda_2$ only. By the AdS/CFT
dictionary, we find
\begin{equation}
\label{eq:jjcorrelator}
\begin{split}
\textrm{Im} \vev{j_i(-p_0)j_j(p_0)}'
&= \frac{\delta_{ij}\, \Omega_{d-2} p_0^{d+2} }
{8(d-1)(4\pi)^{d-2} \epsilon^{d-2} |K_{d/2-1}(ip_0\epsilon)|^2}\\
		&\qquad\cdot
\int_0^1 \ud\tilde{\Lambda}_1 \int_0^{1-\tilde{\Lambda}_1}
\ud\tilde{\Lambda}_2
\; \tilde\Lambda_1\tilde\Lambda_2 \;
H(\tilde\Lambda_1,\tilde\Lambda_2)^{(d-1)/2}\;
	\vert I_{\nu, f}(i,\, \tilde\Lambda_1,\,
\tilde\Lambda_2)\vert^2\,,
\end{split}
\end{equation}
where we defined
\begin{equation}
H(x, y) = \left[1 - (x + y)^2\right]\left[1 -(x - y)^2\right]
\end{equation}
and will refer to it as the ``phase volume'' factor.
The correlator $\vev{j_0(-p_0)j_0(p_0)}$, on the other hand,
is killed off by the projector $P_{\mu\nu}(p)$ for the particular
external momentum $(p_0, 0,\dots,0)$ we have chosen here.

As a consistency check, consider the simple case where the boundary
term is set to zero i.e. $f=0$, where conformal symmetry is expected
 to be preserved. In that case the $p_0$  dependence of $I_{\nu, f}$
 can be completely taken out, which is simply $I_{\mu, 0} \sim p^{-1-d/2}$.
 Using also the fact that
 $\lim_{\epsilon \to 0}K_{\frac{d}{2}-1}(\epsilon p )
 \sim (\epsilon p)^{-\frac{d}{2}-1}$, the correlator is therefore
 given by,
\be
\vev{j_i(-p_0)j_j(p_0)}' = \delta_{ij} \zeta p_0^{d-2},
\ee
where $\zeta$ is a $p$ independent constant obtained from the rest
of the integral. The $p_0$ dependence of the correlator is exactly what
is expected from the conformal dimension of a vector
 current in a CFT in $d$-dimensions.

In the case where $f$ is non-zero, the simple scaling behaviour is
disturbed and conformal invariance broken. It is manifest that
in this case the external
momentum $p$ appears as an additional parameter in the loop integral.

We make a comment regarding the loop diagram~(b) in
figure (\ref{wittendiagram}). It is
necessary to include this loop to recover gauge invariance in the bulk.
But the external momentum in this loop factors out completely
and it does not carry any imaginary part, as in the
case in flat-space. To extract the
imaginary part this diagram can be safely ignored, and its
real part can be inferred via gauge invariance from diagrm~(a).
For this reason, we will omit this loop diagram
in what follows.

The picture outlined above is mostly standard. One is easily
tempted to assume that the loop integral can be evaluated
without any obstacles.  There, however, remains a quite serious difficulty.
As we have already alluded to, it turns out that
the ``vertex function'' $I_{\nu, f_1, f_2}$ is not a regular
function for all values of $\Lambda_1$ and $\Lambda_2$. It contains various
divergences of different orders. Expressed in the $2$-dimensional space
spanned by $\{\Lambda_1,\Lambda_2\}$ the region where $I_{\nu, f_1, f_2}$
becomes singular is not isolated but form boundaries, and it diverges
sufficiently fast near those boundaries that it can cause the loop
integral to diverge, even after the vanishing of the ``phase volume''
factor $H^{(d-1)/2}$ in this limit is taken into
account. This divergences is a new type of divergence, different
from the usual UV divergence which still comes about from the
remaining $\ud^dk$ integral. To compute
the imaginary part of the 1-loop correction and obtain a physical
answer, one must first understand the singularity structures of
$I_{\nu, f_1, f_2}$ and regularize the loop accordingly. This is the
topic we turn to in the next subsection.

\subsection{The divergence of the vertex function and the 1-loop integral}
\label{sec:divergence}

Some discussions on the divergence properties of the ``vertex function''
$I_{\nu, f_1, f_2}$ are in order. We have mentioned that as
$f=0, \infty$, the $p$ dependence of $I_{\nu, 0}$ or $I_{\nu, \infty}$
can be completely taken out and apart from an overall power-law
dependence of $p$, the rest, including
any possible poles of course, depends on the ratio
$\tilde\Lambda_{1,2}\equiv\Lambda_{1,2}/p$ only.  When $f_{1,2}\ne 0$,
there is no longer such a simple scaling behavior of
$I_{\nu, f_1, f_2}$. However,
as it's explained later in this subsection as well as in
Appendix \ref{app:sec3_details},
as far as the divergence of $I_{\nu, f_1, f_2}$
is concerned, a simple scaling behavior still arises and
its singularity structures depend on the ratios $\tilde\Lambda_{1,2}$
and $\tilde f\equiv f/p$ only.
Therefore, without loss of generality, we will set $p=1$ and
assume $\tilde\Lambda_{1,2}=\Lambda_{1,2}$ in this subsection,
ignoring the scaling of $f$ whenever it's unimportant.

We will outline the main point and provide the essential results
in this section, and leave the full details in Appendix
\ref{app:sec3_details} for
those who are interested.

Impressively, Bailey \cite{bailey} evaluated the following integral
involving a power and three Bessel's functions and found
an analytic answer:
\begin{equation}
\label{eq:bailey}
\begin{split}
\int dz z^{\lambda -1} &K_\rho (c z)J_\mu(a z) J_\nu(b z) =
\frac{2^{\lambda-2}a^\mu b^\nu\Gamma[1/2(\lambda + \mu + \nu + \rho)]
	\Gamma[1/2(\lambda + \mu + \nu - \rho)]}{c^{\lambda+ \mu +\nu}
		\Gamma[ \mu + 1]\Gamma[ \nu + 1]}  \nonumber \\
&\qquad\cdot F_4[\frac{1}{2}(\lambda + \mu + \nu - \rho);
\frac{1}{2}(\lambda + \mu + \nu + \rho);\mu+1;\nu +1; -a^2/c^2;-b^2/c^2]\,.
\end{split}
\end{equation}
Here, $F_4$ is the Appell hypergeometric function, one of the
generalized hypergeometric function that, apart from four parameters
fixed in the current case by the coefficients $\lambda, \rho, \mu$ and $\nu$,
depends on two complex variables. For the loop integral that we wish to
compute in the Lorentzian signature, we need to analytically continue Bailey's
result, substituting $c$ by $i c$, and obtain the following identity:
\begin{equation}
\label{eq:continued_bailey}
\begin{split}
I\equiv &\int_0^{+\infty}
z^{\frac{d}{2}} K_{\frac{d}{2}-1}(iz)
J_\mu(\Lambda_1 z) J_\nu(\Lambda_2 z)\ud z\\
	=&
\frac{\Gamma\left[1+\frac{1}{2}(\mu+\nu)\right]\Gamma
	\left[\frac{1}{2}(d+\mu+\nu)\right]\Lambda_1^\mu\Lambda_2^\nu}
{(i)^{d/2+1+\mu+\nu}\Gamma(\mu+1)\Gamma(\nu+1)}\,
	F_4[1+\frac{1}{2}(\mu+\nu),\frac{1}{2}(d+\mu+\nu);\, \mu+1, \nu+1;\,
\Lambda_1^2, \Lambda_2^2]\,.
\end{split}
\end{equation}
This equation, obtained by analytically continuing the
convergent integral \eqref{eq:bailey}, agrees
with the regularization scheme for $I_{\nu, f_1, f_2}$ explained
earlier where one endows $p$ with a small positive imaginary part that
is taken to zero in the end.
Within the region $|\Lambda_1|+|\Lambda_2|<1$, the Appell
hypergeometric function has a series expansion
which fails to converge as $|\Lambda_1|+|\Lambda_2|\rightarrow 1$.
A simple observation of the asymptotics of the integrand
in \eqref{eq:continued_bailey} when $z\rightarrow \infty$
leads to the same conditions for $I$ to be well defined.
In the current context, both $\Lambda_{1,2}$ are real and positive,
which allows us to consider them as the mass of the KK-scalars,
and the condition $\Lambda_1+\Lambda_2\le 1$ appears to be
nothing other than the consequence of energy-momentum conservation,
i.e., a particle can only ``decay'' into two scalars whose masses
add up to a value smaller than the proper energy of the decaying
particle.  But the fact that $I$ approaches infinity
as $\Lambda_1+\Lambda_2\rightarrow 1$ is somewhat problematic and
a regularization scheme is needed.

To this end, we need to understand exactly how the Appell hypergeometric
function $F_4$ diverges as $\Lambda_{1,2}$ approaches
the convergence boundary.
It is sufficient to investigate the asymptotics of the integrand in
\eqref{eq:continued_bailey}
as far as the singularities are concerned. We assume
that the values of $d$, $\nu$ and $\mu$ are chosen properly such
that the integrand is regular at any finite value of $z$, or if it
does contain any singularity at $z<+\infty$, it is a singularity
that can be integrated across and leads to no singular behavior
of $I$.  Given such assumptions, the only possible source for $I$
to diverge is when the integral is carried out all
the way toward $z\rightarrow +\infty$.
In the region that $z$ is large, we can approximate
the Bessel functions by their asymptotic expansions and formally
write
\begin{equation}
\label{eq:asymptotic_I}
I\sim\frac{1}{2^{\frac{d}{2}}\sqrt{\pi\Lambda_1\Lambda_2}}
\sum_{n=0}^\infty\sum_{k=0}^n\sum_{l=0}^{n-k}\int_0^{\infty}
\ud z\,
\frac{\left(\frac{d}{2}-1, n-k-l\right)(\mu, k)(\nu, l)}
{(2z)^{n+\frac{3-d}{2}}\Lambda_1^k\Lambda_2^l}
\left(e^{-i(\theta_\mu+\theta_\nu+\frac{3\pi}{4})}i^{2(k+l)-n}
	e^{-i[1-(\Lambda_1+\Lambda_2)]z}+\dots\right)\,,
\end{equation}
where  $\theta_\mu\equiv\frac{\mu\pi}{2}$ and
the ellipsis represents similar terms that involve other
combinations of $1\pm\Lambda_1\pm\Lambda_2$, whose full form
is given in \eqref{eq:full_asymp_I}. The notation $(\nu, n)$ is
defined as
\begin{equation}
(\nu, n)\equiv\frac{\Gamma(\frac{1}{2}+\nu+n)}
{n! \Gamma(\frac{1}{2}+\nu-n)}\,.
\end{equation}
If we just carry out the integration term by term,
regularizing the oscillatory integrand
by lifting the integral contour slightly above the real axis as
explained earlier, we obtain:
\begin{equation}
\label{eq:sing_I}
\begin{split}
I\sim\frac{1}{2^{\frac{3}{2}}\sqrt{\pi\Lambda_1\Lambda_2}}
\sum_{n=0}^\infty\sum_{k=0}^n\sum_{l=0}^{n-k} &
\frac{\left(\frac{d}{2}-1, n-k-l\right)(\mu, k)(\nu, l)
\Gamma(\frac{d-1}{2}-n)}
{2^n\Lambda_1^k\Lambda_2^l}\\
&\cdot
\left(e^{-i(\theta_\mu+\theta_\nu+\frac{(d-2)\pi}{4})}(-)^{k+l}
		[1-(\Lambda_1+\Lambda_2)]^{n+\frac{1-d}{2}}
+ \dots\right)\,.
\end{split}
\end{equation}
Again, the ellipsis represents similar terms that involve
other combinations of $1\pm\Lambda_1\pm\Lambda_2$ and are given in full
in \eqref{eq:full_expan_I_1} and \eqref{eq:full_expan_I_2}.

We must elaborate a bit more on the procedure outlined above.
The two formulae just given would be flawed if
the ``$\sim$'' were taken to be
equal, because the expansions on the r.h.s. of
these ``equations'' are asymptotic
expansions only. At any fixed value of $z$, the infinite sum does not
converge.  Usually, integrating an asymptotic
expansion term by term is only meaningful
if the result is also considered as the asymptotic expansion of
the true integral when the lower integral limit approaches
infinity. However, since for a given integer $N$,
the sum of the first $N$ terms in the expansion
approximates the full integrand with an arbitrarily
small error when $z\rightarrow \infty$,
if one formally integrates the expansion term by
term and drops everything that
remain finite as $|\Lambda_1|+|\Lambda_2|\rightarrow 1$
along any trajectory, only a finite number of singular terms remain
and they describe precisely the same singularities of
the original integral \eqref{eq:continued_bailey}.
Readers may find it suspicious that
integrating the asymptotic expansion
from $z=0$ to $+\infty$ could lead to anything meaningful since
it is only a good approximation to the actual integrand
when $z$ is sufficiently large.  More appropriately, one should
choose a cutoff scale $L$ that is large but
fixed and separate the infinite integral into
two parts: an integral from $0$ to $L$, and an integral from $L$ to
$+\infty$. The first piece necessarily contains no
poles of $\Lambda_{1,2}$
because $L$ is finite and the integrand is regular. The
second piece must consequently include all the singularities
of \eqref{eq:continued_bailey}, which are
independent of the arbitrarily chosen
cutoff scale $L$. \footnote{Take $d=4$ and the leading term in
the asymptotic expansion as an example, the relevant integrals is
\begin{equation}
I_\textrm{leading}=\textrm{finite piece}
+\int_L^\infty\ud z\, \sqrt{z} e^{-i\tilde pz}
=\frac{e^{-iL\tilde p}\sqrt{L}}{i\tilde p}
	+\frac{\sqrt{\pi}\textrm{erf}(\sqrt{iL\tilde p})}{(2i\tilde p)^{3/2}}
	\,,
\end{equation}
where $\tilde p$ stands for any one of the four combinations
$p\pm\Lambda_1\pm\Lambda_2$. To identify the poles of $\tilde p$,
one takes the limit $\tilde p\rightarrow 0$ and finds that the leading
divergence is precisely given by $\sqrt{\pi}/(2i\tilde p)^{3/2}$.
Anything that depends on $L$ remains finite in the
limit $\tilde p\rightarrow 0$.
The apparent additional $1/\tilde p$ term in the above equation is
cancelled out at this order by an identical piece that arises in the expansion
of the error function near the pole.}  Therefore, as far as
the singularities are concerned, one can choose an
arbitrary cutoff $L>0$,
replace the integrand by its asymptotic form, and carry out
the integration from $z=L$ to $\infty$ term by term. The resulting
expansion contains finite number of terms that diverge as
$|\Lambda_1|+|\Lambda_2|\rightarrow 1$.
Evaluated sufficiently close to the poles, these divergent terms
become independent of $L$, and setting
$L=0$ is only the most convenient choice. We must emphasize that
this method would not in general lead to any
useful information regarding the
finite part of the original integral.  Therefore, the ``$\sim$''
sign in equation \eqref{eq:asymptotic_I} means ``equal up to an arbitrary
regular function''.

It's worth noting that exceptional cases do exist when such an
analysis leads to stronger results. In particular,
if in \eqref{eq:continued_bailey} $d$ is odd and $\mu$ and $\nu$ are
half integers, the asymptotic expansions for
Bessel functions of half-integer orders are truncated to
finite sums in which case the ``$\sim$'' can be replaced by the equal sign
and the equations are exact.  Consequently, for $d=3$, one
finds examples where a full analytic
result can be obtained, as presented in section~\ref{sec:exact_example}.

In the conformal limit, we take $\mu=\nu$ and readily find
the singular terms of $I_{\nu,\infty}$ or $I_{\nu, 0}$ from
equation \eqref{eq:sing_I} when
$1-(\Lambda_1+\Lambda_2)\rightarrow 0$.
For general ``mixed'' boundary conditions we consider,
$f\ne0$ and the dependence of $I_{\nu, f_1, f_2}$ on $p$
is more complicated.
But the asymptotic behavior of $J_{\nu, f}$
as $z\rightarrow +\infty$ is quite simple
and essentially identical
to that of standard Bessel functions $J_{\pm\nu}$,
except for a simple
phase shift explained in details in Appendix
\ref{app:sec3_details}. We can therefore follow
the same procedure and easily obtain the most important result
of this section: the singularities of $|I_{\nu, f_1, f_2}|^2$
within the domain $0\le \Lambda_1, \Lambda_2\le 1$ are described by
\begin{equation}\label{Inuf}
\begin{split}
|I_{\nu, f_1, f_2}(i, \Lambda_1, \Lambda_2)|^2\sim
\frac{\Gamma\left(\frac{d-1}{2}\right)^2}
{8\pi\Lambda_1\Lambda_2}
\bigg\{\frac{1}{[1-(\Lambda_1+\Lambda_2)]^{d-1}}& + \left[\frac{d-1}{2}
-\frac{2(\nu, 1)}{d-3}\left(\frac{1}{\Lambda_1}
+\frac{1}{\Lambda_2}\right)\right]
\frac{1}{[1-(\Lambda_1+\Lambda_2)]^{d-2}}\bigg\}\\
&\qquad+O\left([1-(\Lambda_1+\Lambda_2)]^{-(d-1)/2}\right),\qquad
(d>3)\,.
\end{split}
\end{equation}

We've arrived at a pleasant surprise, which is probably physically
well expected: the leading first and the second divergence
of $|I_{\nu, f_1, f_2}(i, \Lambda_1, \Lambda_2)|^2$
are independent of the parameter $f$ that describes the mixed
boundary condition for the scalar
field $\phi$. The leading divergence is furthermore independent of
its bulk mass parameterized by $\nu$. The remaining singular terms,
on the other hand, do in general depend both on $f$ and $\nu$. But
fortunately, the orders of those ``true'' singularities are equal or less
than $(d-1)/2$, just sufficiently low to be suppressed by the phase volume
factor
\begin{equation}
H^{(d-1)/2}
=\left\{[1-(\Lambda_1+\Lambda_2)]\,[1+\Lambda_1+\Lambda_2]\,
[1-(\Lambda_1-\Lambda_2)]\,
[1-(\Lambda_2-\Lambda_1)]\right\}^{(d-1)/2}\,,
\end{equation}
which contains precisely all four combinations of
$(1\pm\Lambda_1\pm\Lambda_2)$
to the power of $(d-1)/2$.  Hence, the $f$-dependent imaginary part of
the 1-loop integral \eqref{eq:jjcorrelator} is finite and can be
evaluated free of any pathologies.  This conclusion holds true
for any $d\ge 2$.

It should be noted once again that, in \eqref{eq:sing_I}, only terms that
become singular as $1-(\Lambda_1+\Lambda_2)\rightarrow 0$
should be kept and the rest must be ignored, unless
the asymptotic expansion is known to converge.
For example, when $d<4$, only the leading divergent term in \eqref{eq:sing_I}
is really there.
The $1/[1-(\Lambda_1+\Lambda_2)]^{(d-3)/2}$ term and therefore the
second term in expansion \eqref{Inuf} do not exist.
Similarly, terms as $1/(1-\Lambda_1-\Lambda_2)^{(d-5)/2}$
in \eqref{eq:sing_I} is to be discarded if $d<6$.
Precisely as the method fails, the coefficients in expansion \eqref{Inuf}
become singular.

We remind the readers that for general values of $\Lambda_{1,2}$,
$I_{\nu, f_1, f_2}$ contains other singularities whenever
$1\pm\Lambda_1\pm\Lambda_2\rightarrow 0$ and an analogous expansion
can be obtained in the same manner near each of them
as shown in Appendix \ref{app:sec3_details} \footnote{Although
one needs to extend the definition for $I_{\nu, f_1, f_2}$
outside $0\le\Lambda_1+\Lambda_2\le 1$ by choosing the
integral contour for $z$ with special care.}. In fact there's a
natural way to understand why all those singularities are essentially
the same. We take this chance to mention some other interesting
properties of the vertex function which might be useful for a full
calculation of 1-loop integrals in AdS space. For brevity, we will
ignore the differences between photon and scalar and consider
a bulk to bulk 1-loop integral \cite{in_progress}, in which the
following vertex function appears:
\begin{equation}
\label{eq:JJJ}
I(p, \Lambda_1, \Lambda_2)=
\int_0^\infty \ud z \; z^\alpha J_\nu(pz) J_\nu(\Lambda_1 z)
	J_\nu(\Lambda_2 z)\,,
\end{equation}
where $\alpha$ and $\nu$ depend on the details of the theory.
We'd like to point out that if one can ignore the subtleties
related to the singular behavior of the integrand
at $z\rightarrow \infty$, formally this function $I$ is defined on
$\mathbb{CP}^2$ if all variables are complex numbers, apart from a
trivial power-law factor.  This is because one
can always absorb an overall scaling of $(p, \Lambda_1, \Lambda_2)\rightarrow
(\lambda p, \lambda \Lambda_1,\lambda \Lambda_2)$ by redefining
$z\rightarrow \lambda^{-1} z$
together with a change of integral contour which always connects
$0$ to $\infty$ in the complex plane. If we restrict ourselves to
the real space, one can only absorb such a scaling if $\lambda>0$, so
$I$ is defined on $S^2$. Let us focus on the case when $p$
and $\Lambda_{1,2}$,
are real and positive. A simple scaling of $z$ immediately allows us to
set $p=1$.  We find there are four different regions in the first quadrants
in the $(\Lambda_1, \Lambda_2)$ plane separated from each other
by singular boundaries.  They are
$\textrm{I}=\{0\le\Lambda_1+\Lambda_2\le 1\}$,
$\textrm{II}=\{1\le \Lambda_1-\Lambda_2\}$,
$\textrm{III}=\{1\le \Lambda_2-\Lambda_1\}$, and
$\textrm{IV}=\{1\le\Lambda_1+\Lambda_2\,,\,
	|\Lambda_1-\Lambda_2|\le 1\}$.
We have been focusing on region $\textrm{I}$ only in the
above discussion. But
region $\textrm{II}$ and $\textrm{III}$ can be mapped to $\textrm{I}$ by
a simple scaling. To go from region II to region I, we just
scale $z\rightarrow \Lambda_1 z$ in \eqref{eq:JJJ} and define
$\tilde\Lambda_2=\Lambda_2/\Lambda_1$, and
$\tilde\Lambda_1=1/\Lambda_1$. It's easily verified that
$0\le\tilde\Lambda_1+\tilde\Lambda_2\le 1$, and
\begin{equation}
I(1, \Lambda_1, \Lambda_2)
=\tilde \Lambda_1^{\alpha+1} I(1, \tilde\Lambda_1, \tilde\Lambda_2)\,.
\end{equation}
Similarly, one can go from region $\textrm{III}$ to region $\textrm{I}$
by a simple scaling of $z\rightarrow \Lambda_2 z$, defining
$\tilde\Lambda_2=1/\Lambda_2$ and $\tilde\Lambda_1=\Lambda_1/\Lambda_2$,
and verifying that $I(1, \Lambda_1, \Lambda_2)
=\tilde \Lambda_2^{\alpha+1} I(1, \tilde\Lambda_1, \tilde\Lambda_2)$.
Notice that $I(1,\Lambda_1,\Lambda_2)$ actually vanishes quite
fast as $\Lambda_{1,2}$ is large.
Region IV is more complicated, but if one is only interested in
the asymptotics of the integrand in \eqref{eq:JJJ}, which dictates
the singularities of $I$ as we argued above,
one can also map it to region I by a scaling
$z\rightarrow (\Lambda_1+\Lambda_2) z$,
define $\tilde \Lambda_1=(1+\Lambda_1-\Lambda_2)/[2(\Lambda_1+\Lambda_2)]$
and $\tilde\Lambda_2=(1+\Lambda_2-\Lambda_1)/[2(\Lambda_1+\Lambda_2)]$,
and find that the singularities of $I(1,\Lambda_1, \Lambda_2)$ are the
same as those of $I(1,\tilde\Lambda_1, \tilde\Lambda_2)$ but with
$(\tilde\Lambda_1,\tilde\Lambda_2)$ located in region I now.
Therefore all the singularities of $I$
as $|\Lambda_1|+|\Lambda_2|\rightarrow 1$ are related to each other in
a simple way.

\begin{figure}[ht]
\includegraphics[width=0.58 \textwidth]{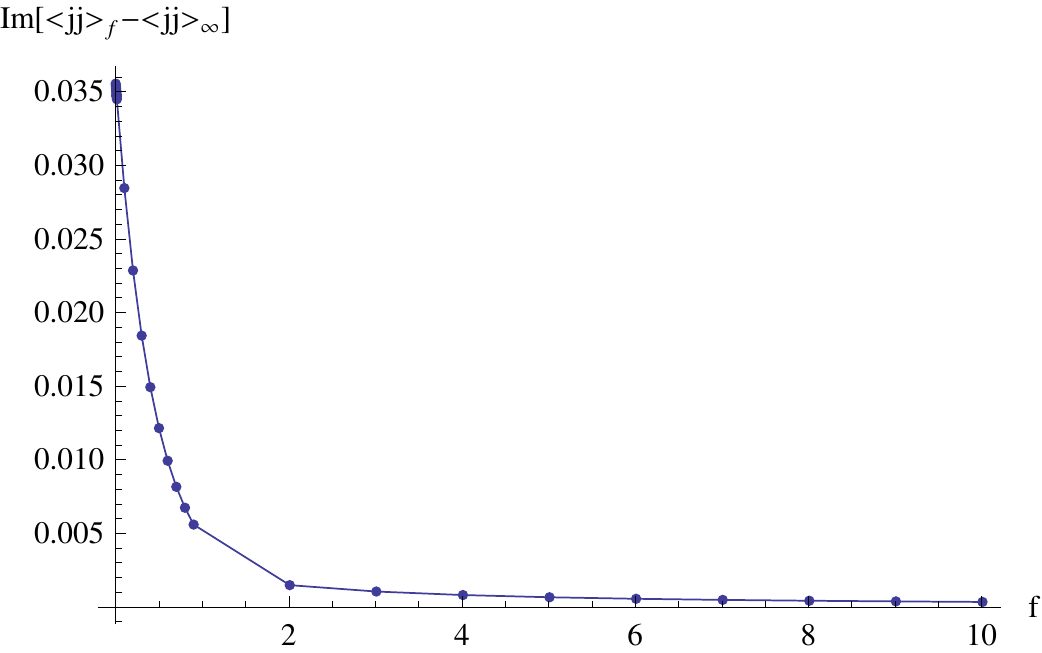}
\caption{The relative variation of the imaginary part of
	the 1-loop integral against coupling $f$ at $d=4, \nu=\frac{1}{2}$.
	External momentum being an overall scale is set to one. }
\label{jjplot}
\end{figure}

Just as a demonstration, we show in figure (\ref{jjplot}) the
numerical result for the relative
variation of the imaginary part of the 1-loop diagram against
the boundary coupling $f$ when $d=4$ and $\nu=\frac{1}{2}$. The numerical
result presented in the plot is subtracted against that when $f=\infty$.

\subsection{A few more general observations on the
divergences in AdS loops}
Thus far we have studied closely loops involving charged scalars
coupled to photons and the divergences associated to the vertex.
We would like to make a few observations about loops in AdS
more generally. We found
that this kind of divergence occurring as the vertices approach
the AdS horizon is fairly general.  Consider for simplicity a $\phi^n$
vertex of the form
\be
\mathcal{V}(\phi)=\int \prod_i^n \ud^d p_i \int_0^\infty \ud z
\sqrt{-g}\, \delta^d\left(\sum_i^n p^\mu_i\right)\prod_i^n \phi(p_i,z)
\ee
expressed in momentum space for the flat directions. Substituting in
the integral the radial wave-function of the scalar field of mass $m$,
which, for Neumann boundary condition as an example, is given by
$z^{d/2}J_{\nu}(p z)$, where $p=|p^\mu|$, and taking the near horizon limit,
we find
\be
\mathcal{V}(\phi) \sim \int \prod_i^n \ud^d p
\delta^d\left(\sum_i^n p^\mu_i\right)
	\int^\infty \ud z  (z^{-d-1}) z^{n\frac{d-2}{2}}
	\sum_{\varepsilon_i=\pm1}\cos\left[
\sum_i^n(\varepsilon_i p_i z+\varepsilon_i\theta_\nu)\right]\,.
\ee
We have used the asymptotic behavior
$\lim_{z\to\infty} J_{\nu}(p z) \sim \frac{1}{\sqrt{p z}}
\cos(pz+\theta_\nu)$.
By naive power counting, ignoring the oscillatory factors, this
integral is only convergent if
\be
\frac{n(d-2)}{2}< d+1\,,
\ee
precisely identical to the renormalizability condition one would find
via naive counting of engineering dimensions. Notice this analysis
is independent of the boundary condition for the scalar since
it would only affect the oscillatory factors and leave the
power-law dependence intact.

Such a coincidence no longer exists as soon as we consider coupling to
fields with higher spins. In the case of photons, the vertex would
appear divergent by the above naive power-counting for arbitrary
$d$ independently of the renormalizable conditions,
because the coupling invariably involves the inverse metric which
provides extra factors of $z^2$ even though the photon
propagator is proportional only to $z^{\frac{d-2}{2}}J_{{d/2-1}}$,
and subsequently contribute to the divergence in the horizon limit.
This consideration is generally applicable to the coupling of scalars and
graviton as well.

It is very important to note that even as the $z$-integral
appears divergent according to
the above consideration, it does not, however, necessarily imply that the
loop integral is divergent.  The vertex integral can be regularized
by analytic continuation as we have done in our earlier
discussions making use of its oscillatory
factors. Using the method introduced in
the previous section, one can easily verify that this vertex after
being regularized must include poles at $\sum_i^n \varepsilon_i p_i=0$,
where $\varepsilon_i=\pm 1$, and could potentially lead to similar
divergences in loop integrals in the collinear limit
as $\sum_i^n p_i\rightarrow 0$.
Depending on the order of these poles, which must be analyzed case by case
following the procedure explained in section \ref{sec:divergence},
this may or may not lead to pathologies.
Let us mention, as an example here, the same analysis
implies that this extra divergence occurring as the vertex in
diagram (b) in figure (\ref{wittendiagram}) approaches the
horizon is logarithmic if $d=4$.

We would also like to point out here that using the
form of fermion wave-function in AdS as given in
\cite{Iqbal:2009fd}, we can repeat the above power counting
for a fermion-photon vertex as $z\to \infty$ and find
\be
\int^\infty dz  \sqrt{-g}A_\mu e^{\mu}_a \bar{\psi}
\Gamma^a \psi \sim \int^\infty dz z^{\frac{d-3}{2}}
\times\textrm{oscillatory factors}\,,
\ee
which is exactly the same as in the case of scalars.
Here $e^{\mu}_a$ is a simple diagonal choice for the vierbeins,
and $\Gamma^a$ are flat space gamma matrices. It seems possible
that in supersymmetric theories these type of divergences are cancelled
automatically among fermions and bosons.

Finally, we make a passing comment about potential UV
divergence near the boundary $z\to 0$. We have restricted
our attention to scalar fields of mass $m$ such that
$\nu = \sqrt{\frac{d^2}{4}+m^2} \le \frac{1}{2}$ because
this is a window of consistent masses such that UV divergence
near the boundary is simply non-existent, and curiously
this condition is dimensional independent.

It would be interesting to have a better understanding
of the physics behind these observations.

\section{Random disorder and the replica trick in AdS/CFT}
\label{sec:disorder}

We now turn to an interesting application of the results
given in the previous
section and calculate the effect of random disorder
on the transport coefficients using AdS/CFT correspondence.
\subsection{The replica trick and the conductivities}
We begin with a brief review of the replica trick.
Consider a quantum system described by the action $S$.
The effects of random disorder can be captured by
introducing into the action an extra scattering potential $\delta S$
\be
\delta S = \int d^d x V(x,t) \mathcal{O}(x,t),
\ee
where $V(x,t)$ is a general spacetime dependent random
scattering potential, and $\mathcal{O}$ is some physical operator, such as
the charge density etc. In the presence of the random
scattering potential term, correlation functions
averaged over the random variable $V$ is given
by
\begin{equation}
\overline{\langle \mathcal{O}(x_1)\mathcal{O}(x_2)... \rangle}
= \int D[V] P[V] \left( \frac{\int D\phi e^{-S- \delta S}
\mathcal{O}(x_1)\mathcal{O}(x_2)...}{\int D\phi
e^{-S_0- \delta S} } \right),
\end{equation}
where $P[V]$ is the probability distribution of the random
potential and
the over-line denotes averaging over $V$. A well-known trick in the
condensed matter literature to compute this averaged correlator
is called the \emph{replica trick}.
The idea is to introduce $n$ copies of the theory concerned such that
all operators in the theory are replicated $n$ times and we label
them by an extra index $i\in\{1,...,n\}$.
The partition function  of the full theory including the $n$ copies
is related to that of the original one by $\mathcal{Z}_n = (\mathcal{Z}_1)^n$.
Then the correlation function for a fixed $V$ in the original theory
is formally related to the $n$-replica partition function, treating
$n$ as a continuous variable by analytic continuation as
\begin{equation}
\langle \mathcal{O}\mathcal{O}...\rangle_V = -\lim_{n\to 0}
\left.\frac{\delta}{\partial J_1}\frac{\delta}{\partial J_1
}...\frac{1}{n} (e^{n\ln \mathcal{Z}_1}-1)\right|_{J_i=0}
=-\lim_{n\to 0}\left. \frac{\delta}{\partial J_1}\frac{\delta}
{\partial J_1}...\frac{1}{n} \mathcal{Z}_n\right|_{J_i=0}\,,
\end{equation}
where $J_i$ is the current coupled to the replicated operator
$\mathcal{O}_i$. The averaged correlation functions can be readily
evaluated as
\begin{equation}
\label{eq:replica_master}
\overline{\vev{\mathcal{O}\mathcal{O}\dots}}= -\lim_{n\to 0}
\left.\frac{\delta}{\partial J_1}\frac{\delta}{\partial J_1}...
\frac{1}{n} \int D[V] \,e^{\ln P[V]}\,\mathcal{Z}_n
\right|_{J_i=0}= -\lim_{n\to 0}\
\left.\frac{\delta}{\partial J_1}\frac{\delta}{\partial J_1}...
\frac{1}{n} \,\tilde{\mathcal{Z}}_n\right|_{J_i=0}\,.
\end{equation}
We have exchanged the order of taking the $n\rightarrow 0$ limit
and the integral of $V$ in the first step and defined
\begin{equation}
\tilde{\mathcal{Z}}_n\equiv\int D[V]\prod_i D[\phi_i]
e^{-\sum_{i=1}^n (S_i+\delta S_i)+\ln P[V]}
\end{equation}
in the second. For more complete explanation,
see \cite{Cardy} and references therein.

This relation \eqref{eq:replica_master} is useful
because in the simple but common situation where the
random distribution $P[V]$ is simply Gaussian and local
in space and time i.e.
 \begin{equation}
P[V] = e^{- \frac{V^2(x,t)}{2 f}}\,,
 \end{equation}
the scattering potential $V$ can be readily integrated out, and we have
\begin{equation}
\tilde Z_n=\int \prod_i D[\phi_i] e^{-\sum_{i=1}^n S_i
	-\frac{f}{2}\int d^dx \sum_{ij}\mathcal O^i \mathcal O^j}.
\end{equation}

To study the effects of random disorder in a strongly coupled theory, it
is natural to incorporate the replica trick in the context of AdS/CFT
correspondence, which is considered in \cite{Hartnoll:2008hs,Fujita:2008rs}.
The spirit of the two studies are very different and
in this section we follow \cite{Fujita:2008rs} and attempt
to study the effects of disorder on the conductivity
of a general charged system.

In \cite{Fujita:2008rs} the replicas are introduced
literally as $n$ copies of the AdS backgrounds. The
random scattering potential in the CFT
in the replicated theory is presented by the boundary term
\be
\delta S = \int_{\partial \mathcal{M}_{d+1}}
d^dx V(x,t)\sum_i \mathcal{O}^i,
\ee
which gives
\be \label{bterms2}
\delta S = - \frac{f}{2} \int_{\partial
\mathcal{M}_{d+1}} d^dx \sum_{i,j} \mathcal{O}^i \mathcal{O}^j,
\ee
after the potential is integrated out.
These boundary terms are precisely those multi-trace
perturbations in the dual CFT theory discussed in
earlier sections, and dictate the boundary
conditions of the corresponding scalar fields in
the multiple AdS bulks.  They also relate
the boundary values of the scalars in different
AdS copies and thus connect
them together.
The modified boundary-boundary and bulk-to-bulk
propagators of these scalar fields have been studied
in \cite{Gubser:2002zh} and, in the context of the
replica trick, in \cite{Fujita:2008rs}. It is known
that the effect of these boundary perturbations are mediated to other
sectors beginning only at 1-loop order
\cite{Aharony:2001pa,Aharony:2006hz,Kiritsis:2006hy}.
Using the \emph{folding trick} that treats these replicated
bosons as if they live in the same AdS space \cite{Aharony:2006hz}
we can thus apply techniques discussed in the previous
sections to compute the
conductivities in the presence of random disorder.

Now consider again scalar field $\phi$ with mass $m$.
Recall that its boundary expansion is given by
\be
\phi \sim \alpha z^{\Delta_+} + \beta z^{\Delta_-},
\qquad \Delta_\pm = \frac{d}{2}
\pm \sqrt{m^2 + \frac{d^2}{4}}. \nonumber
\ee

In the presence of the boundary deformations (\ref{bterms2}),
the scalar fields concerned satisfy nontrivial
boundary conditions.
\begin{equation}
\alpha_i = f \sum_j^n \beta_j,
\end{equation}
where $n$ is the number of replicas introduced.
These conditions can be diagonalized by:
\be\label{diagbasis}
\tilde{\phi}_l = \sum^n_j a^{l}_j \phi_j,
\qquad 0<l<n, \qquad a^{n}_j =\frac{1}{\sqrt{n}},
\qquad \sum_j a^{l\ne n}_j = 0\,.
\ee
There is only one linear combination of the
fields $\phi_i$
(corresponding to $\tilde{\phi}_n$ defined above) satisfies
the mixed boundary conditions depending on the
strength of the disorder $f$, i.e.,
\be \tilde{\alpha}_{n} = n f \tilde{\beta}_{n}, \label{tbc1}
\ee
whereas the rest of the fields $\tilde{\phi}_{l\ne n}$
simply satisfy usual Neumann boundary conditions
\be \tilde{\alpha}_{m \ne n} =0. \label{tbc2}\ee

It would be useful to pick an orthonormal basis for
these fields, so that the bulk-to-bulk propagators
take a simple form.
One convenient choice is
\be
a^{(n-l)}_i =\frac{1}{\sqrt{(n-l)(n-l+1)}}
\bigg\{ \begin{array}{cc} 1, & 1\le i\le (n-l)\\
-(n-l),& i= n-l+1 \\
0 ,     & i> n-l+1 \,. \end{array}
\ee

The bulk-to-bulk propagators of the fields $\phi_i$
in the original basis would be related to these
rotated basis $\tilde{\phi}$ by simple linear
combinations. In particular, we are interested
in the following propagators
\begin{eqnarray}\label{propmix}
&&G^{nn}= \frac{1}{n} (\tilde{G}^{nn}+(n-1)
\tilde{G}^{(n-1)(n-1)}),\qquad G^{i j}=
\langle \phi_i \phi_j\rangle,\qquad
\tilde{G}^{i j}= \langle \tilde{\phi}_{i}
\tilde{\phi}_{j}\rangle, \nonumber \\
&&G^{n(n-1)}= \frac{1}{n}
(\tilde{G}^{nn}-\tilde{G}^{(n-1)(n-1)})\,.
\end{eqnarray}

The bulk-to-bulk propagators of the orthonormal
scalar fields $\tilde{\phi}_l$ satisfy the usual
Klein-Gordon equations in $AdS_{d+1}$ space, as
discussed in the previous sections.  Explicitly,
we have, in Euclidean signature and in the $n\to 0$ limit,
\begin{equation}\label{nnprop}
\begin{split}
G^{i i}(x,y,k) &=\int d\Lambda \frac{\Lambda}
{\Lambda^2 + k^2}\bigg( J_{-\nu}(\Lambda x)J_{-\nu}
(\Lambda y) + \frac{f\Gamma(1+\nu)}
{\Gamma(1-\nu)(2\Lambda)^{2\nu}}
(J_{\nu}(\Lambda x)J_{-\nu}(\Lambda y)\\
&+ J_{-\nu}(\Lambda x)J_{\nu}(\Lambda y)
- 2\cos(\nu \pi)J_{-\nu}(\Lambda x)J_{-\nu}
(\Lambda y))\bigg),
\end{split}
\end{equation}
and
\begin{equation}
G^{i,j\ne i}(x,y,k) =\int \frac{\Lambda\,\ud\Lambda}
{\Lambda^2 + k^2}\left[
\frac{f\Gamma(1+\nu)}{\Gamma(1-\nu)(2\Lambda)^{2\nu}}
(J_{\nu}(\Lambda x)J_{-\nu}(\Lambda y)+ J_{-\nu}
(\Lambda x)J_{\nu}(\Lambda y)-
2\cos(\nu \pi)J_{-\nu}(\Lambda x)J_{-\nu}(\Lambda y))\right].
\end{equation}

The conductance in Lorentzian signature, related to
$\overline{\langle j_i(-p)j_j(p)\rangle}$ and
$\overline{\langle j_i(-p)\rangle\langle j_j(p)\rangle}$,
are thus given by precisely the same 1-loop calculation
as detailed in the previous section, except the
scalar bulk-to-bulk propagators are replaced
by $G^{ii}$ and $G^{i,j\ne i}$ respectively.

Given the notable complication in the form of the
propagators (\ref{propmix}), one would like to know
if the divergence occurring in the collinear limit
of the momenta, discussed extensively in the previous
section, could actually acquire some $n$ or boundary
coupling $f$ dependence, rendering our procedure
in extracting the transport properties pathological.
Fortunately, one can show that just as before, the
divergent terms that are not extinguished by the
phase space volume factor are again $n$ and $f$ independent.

To see that, let us rewrite for example the basis rotation:
\be
\phi_n = \frac{1}{\sqrt{n}}
\tilde{\phi}_n - \sqrt{\frac{n-1}{n}}
\tilde{\phi}_{n-1} = \cos \xi \tilde{\phi}_n - \sin\xi \tilde{\phi}_{n-1},
\ee
where we remind our readers that the tilde fields are
those that diagonalize the boundary conditions. We
deliberately rewrite the coefficients in terms of
sine and cosine to make it explicit that the squares of
these coefficients add up to one. The corresponding propagator is
\be
G^{nn} = \cos^2\xi \tilde{G}^{nn} + \sin^2\xi \tilde{G}^{(n-1)(n-1)},
\ee
and to be explicit,
\be
\tilde{G}^{nn} = \int d\Lambda \frac{\Lambda}
{\Lambda^2 + k^2} J_{\nu,n f}J_{\nu,n f},\qquad
\tilde{G}^{(n-1)(n-1)} = \int d\Lambda
\frac{\Lambda}{\Lambda^2 + k^2} J_{\nu,0}J_{\nu,0},
\ee
where $J_{\nu,f}$ are defined in (\ref{besself}). Plugging the
propagator $G^{nn}$ into the loop including the external
photon propagator then leads to the products
\begin{eqnarray}
&&x^{d/2}y^{d/2}K_{\frac{d}{2}-1}(-ipy)K_{\frac{d}{2}-1}(ipx)
\prod_{i=1}^2\left[\cos^2\xi J_{\nu,nf}
(\Lambda_i x)J_{\nu,nf}(\Lambda_i y) + \sin^2\xi J_{\nu,0}
(\Lambda_i x)J_{\nu,0}(\Lambda_i y)\right]\nonumber \\
&&\sim x^{(d-3)/2}y^{(d-3)/2}\exp(i p x)\exp(-ipy)
\prod_{i=1}^2\bigg[\cos^2\xi
(\cos(\Lambda_i x -\theta_{\nu,nf}-\frac{\pi}{4}))
(\cos(\Lambda_i y -\theta_{\nu,nf}-\frac{\pi}{4})) \nonumber \\
&&+ \sin^2\xi (\cos(\Lambda_i x -\theta_{\nu,0}-\frac{\pi}{4}))
(\cos(\Lambda_i y -\theta_{\nu,0}-\frac{\pi}{4}))\bigg],
\end{eqnarray}
where we have omitted some overall numerical factors and
powers of $\Lambda_i$'s which are common to all scalar
loops independently of boundary perturbations.

Concentrating on the collinear limit where one actually
encounters a divergence i.e. $\Lambda_1+\Lambda_2$ is close
to $p$, the dominant contribution from the above
expressions, after doing the radial integral over $x$ and $y$ is given
similar to the discussion leading to (\ref{Inuf}), by
\begin{eqnarray}
&&\sim (p-\Lambda_1-\Lambda_2)^{1-d}\bigg(\cos^4\xi
e^{-i (2\theta_{\nu,nf}-\frac{\pi}{2})}
e^{i (2\theta_{\nu,nf}-\frac{\pi}{2})} + \sin^4\xi
e^{-i (2\theta_{\nu,0}-\frac{\pi}{2})}
e^{i (2\theta_{\nu,0}-\frac{\pi}{2})} \nonumber \\
&&+ 2\cos^2\xi \sin^2\xi
e^{-i(\theta_{\nu,nf}+\theta_{\nu,0} - \frac{\pi}{2} )}
e^{i(\theta_{\nu,nf}+\theta_{\nu,0} - \frac{\pi}{2} )}\bigg)\nonumber \\
&&= (p-\Lambda_1-\Lambda_2)^{1-d}
(\cos^2\xi+ \sin^2\xi)^2 = (p-\Lambda_1-\Lambda_2)^{1-d}.
\end{eqnarray}
The dependence on $n$ and $f$ drops out in the leading divergence.
One could repeat the exercise for the next leading order divergence
to see that again it is $n$ and $f$ independent. Since the expansion
in powers of $n$ is completely regular in the propagators, the
divergences remain $n,f$ independent
even as we expand first in $n$ before doing the integral.
We will in fact work out in detail in the following section an
analytic example where the result after doing the radial
integrals is particularly simple, and explicitly in this
example the divergence can be seen to be $n,f$ independent
as we have demonstrated via the asymptotic expansion method here.

\begin{figure}[ht]
\includegraphics[width=0.58 \textwidth]{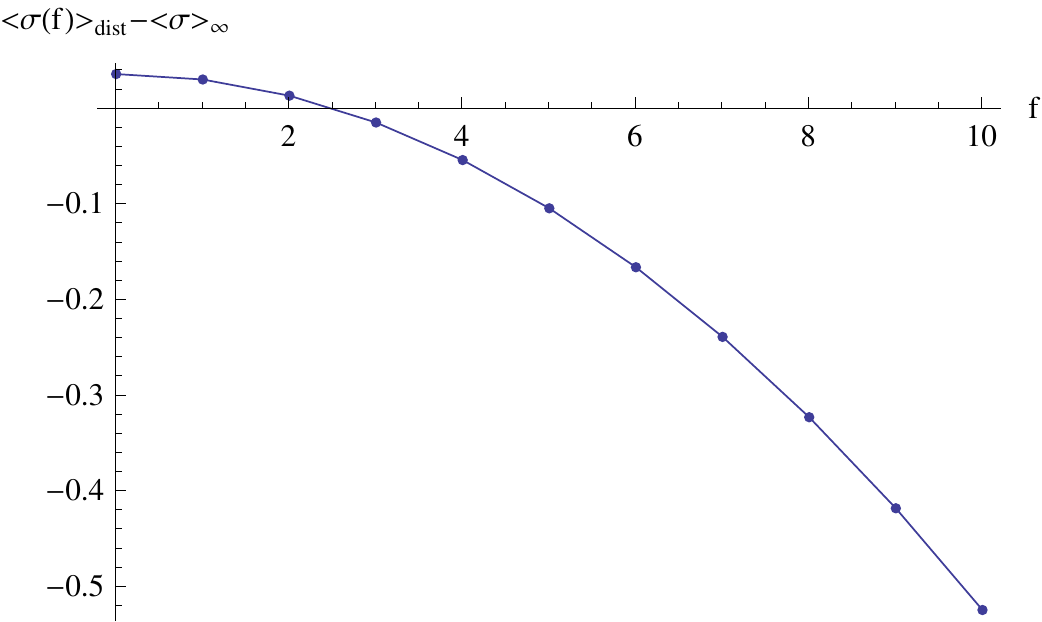}
\caption{Dissipative part of the 1-loop contribution to the
conductivity against coupling $f$ at $d=4, \nu=\frac{1}{2}$.
External momentum being an overall scale is set to one. }
\label{conductdisorder}
\end{figure}

We extract the dissipative part of the conductivity from the
correlation functions via the Kubo formula
\be
\sigma(p_0)=\frac{1}{p_0}\textrm{Im}[\overline{\langle j_{i}
(-p_0)j_j(p_0) \rangle_{\textrm{retarded}}  }].
\ee
We include in figure (\ref{conductdisorder}) a numerical plot
of the conductivity against coupling $f$ at $d=4, \nu =\frac{1}{2}$.
The resultant conductivity is regulated, for convenience, by
subtracting off a corresponding scalar loop satisfying
Dirichlet boundary conditions in a simple un-replicated AdS.
One can see that the conductivity decreases as disorder is
turned on. This point shall be discussed again in our
analytic example in the next section.

We point out that if the random disorder has a non-trivial power function in
momentum space, which corresponds to spatial corelation of these impurities, one can
replace $f$ by an appropriate function $f(p)$ and the results are straightforwardly
generalized by the same procedure.

\subsection{An analytic example in $d=2+1$}
\label{sec:exact_example}
At $d=2+1$ and $\nu = \frac{1}{2}$, the result is particularly
simple. The Euclidean photon propagator given by (\ref{photon_prop})
for $d=3$ reduces to
\be
\label{bulk-boundary}
A_\mu(z,p) = J^{\perp}_\mu(p) \frac{(pz)^{d/2-1}
K_{\frac{1}{2}}(p z)}{(p\epsilon)^{\frac{1}{2}}
K_{\frac{1}{2}}(p \epsilon)},\qquad p^\mu J^{\perp}_\mu(p)=0,
\qquad K_{\frac{1}{2}}(p z) = e^{- p z} \sqrt{\frac{\pi}{2 p z}}\,.
\ee
Similarly, scalar propagators proportional to products of
$J_{\pm \frac{1}{2}}(\Lambda x)$ become simply products of
$\sqrt{2/\pi\Lambda x} \sin \Lambda x$ and
$\sqrt{2/\pi\Lambda x} \cos \Lambda x$.
More explicitly, \eqref{nnprop} becomes in this case
\begin{equation}
\begin{split}
G^{ii}(x,y,k) =& \int d\Lambda\frac{\Lambda}
{\Lambda^2 + k^2} F^{ii}(x,y,\Lambda), \\
F^{ii}(x,y,\Lambda) =&
\bigg(\frac{2 \cos(x \Lambda)
\cos (y \Lambda)}{\pi \Lambda
\sqrt{ x y }}   + \frac{f}{2\pi \Lambda^2\sqrt{
 x y }} (\cos(x \Lambda) \sin (y \Lambda)+
 \sin(x \Lambda) \cos (y \Lambda))\bigg).
\end{split}
\end{equation}

Plugging into the loop integral and denoting the radial coordinates
of the vertices $x,y$, in Lorentzian signature, we have
\begin{eqnarray}
I_{\textrm{disorder}}&=&\int dx dy
K_{\frac{1}{2}}(i p x)K_{\frac{1}{2}}(-i p y)
F^{ii}(x,y,\Lambda_1)F^{ii}(x,y,\Lambda_2)\nonumber \\
 &=& \frac{8 (-1 + \tilde{\Lambda}_1^2 +
 \tilde{\Lambda}_2^2)^2 -
 \tilde{f}^2 (\tilde{\Lambda}_1^4
 + (-1 + \tilde{\Lambda}_2^2)^2 -
    2 \tilde{\Lambda}_1^2
    (1 + \tilde{\Lambda}_2^2))}{4 p^5 \pi \tilde{\Lambda}_1 \
\tilde{\Lambda}_2
(\tilde{\Lambda}_1^4 + (-1 + \tilde{\Lambda}_2^2)^2 -
   2 \tilde{\Lambda}_1^2 (1 + \tilde{\Lambda}_2^2))^2},
\end{eqnarray}
where again we used the dimensionless quantities
\be
\tilde{\Lambda}_i = \frac{\Lambda_i}{p}, \qquad \tilde{f} = \frac{f}{p}.
\ee

It is important to note here that there is again a divergence not
cured by the phase space factor in the collinear limit proportional
to $(p-\Lambda_1-\Lambda_2)^{1-d} =(p-\Lambda_1-\Lambda_2)^{-2}$.
However, it is completely independent of the boundary perturbation
coupling $f$. Given the explicit analytic result of the integral, we
shall subtract off the divergence directly here.

Now we can compute the dissipative part of the conductivity in the presence of the disorder and find
\begin{eqnarray}
\sigma(p_0)_\textrm{1-loop}&&= \frac{1}{p_0}\textrm{Im}
[\overline{\langle j_{i}(-p_0)j_j(p_0) \rangle }_\textrm{1-loop}]
\nonumber \\
&&= \lim_{\epsilon\rightarrow0}\epsilon
\delta_{ij} \frac{(2\pi)^3 p_0^4}{4(4\pi)^3
	|K_{\frac{1}{2}}(ip_0\epsilon)|^2}\int_0^1 d\tilde{\Lambda}_1
\int_0^{1-\tilde{\Lambda}_1}d\tilde{\Lambda}_1
\tilde\Lambda_1\tilde\Lambda_2 H(\tilde\Lambda_1,
\tilde\Lambda_2) I_{\textrm{disorder}} \nonumber \\
&&= \lim_{\epsilon\rightarrow0}
\epsilon\delta_{ij} \frac{- 2 (\frac{f}{p_0})^2
+\frac{925}{72} - \frac{\log(2)}{3} +
\log(8)}{32 p_0 (4\pi)^3 |K_{\frac{1}{2}}(ip_0\epsilon)|^2}.
\end{eqnarray}

The $f$-independent part is simply the 1-loop correction to the
conformal result. The precise value is unimportant, since it
depends on how we subtracted the divergence. The
$f$-dependent contribution however is interesting. First of all,
it is negative definite, which means it always reduces the
conductivity. Secondly, while the loop correction is only of
order $N^0$, it is clear that in the low frequency limit
$(p_0/f^2)^2 \ll N^{-2}$ and this loop correction would eventually
become more important than the tree-level contribution, which is
of order $N^2$. This is not surprising since the boundary perturbation
is deliberately chosen to be relevant in the infrared. This means
that to extract result in the deep infrared limit one probably needs
to re-sum the loop contributions. This could be the explanation as
to how the delta-function at $p_0=0$ is corrected even though
naively the boundary perturbation only begins to contribute at
1-loop level and appears extremely weak. We would like to pursue
this in more detail in future work.\\
\vspace{0.5cm}\\
\underline{\bf Time independent disorder}\\
When the disorder is time independent, only the component of the
scalar $\phi(k,z)$ with $k= (0,k_i)$ acquires a nontrivial
boundary condition (\ref{bc1}). i.e. the disorder coupling $f$ is
frequency dependent and is more precisely given by $f \delta_{p_0,0}$.
The disorder therefore only contributes non-trivially if at least
one of the two internal bulk-to-bulk propagators is evaluated
at exactly zero energy. By
the Cutskovsky rules, the imaginary part of the loop is only
non-zero when $\Lambda_{1,2}$ can satisfy the ``on-shell'' conditions that
restrict them to be real and non-negative.
All these conditions can be satisfied only at one point of the
phase space where $k_i=\Lambda_1=0$. In other words, by putting
the KK-scalar fields on-shell and requiring that one of them
has zero energy, its corresponding spatial momentum and
effective mass has to be zero. Note however that the boundary
disorder coupling $f$ always come in the dimensionless
combination $(\Lambda)^{2\nu}/f$. Therefore as soon as
we set $\Lambda=0$ such that the momentum dependent
coupling $f$ contributes, its contribution is immediately
killed by the $(\Lambda)^{2\nu}$ factor, for any
positive $\nu$. The time-independent disorder therefore
doesn't seem to have any effect on the dissipative
part of the conductivity, even though the operator is
relevant. The only possibility of a non-trivial
contribution seems to be the marginal limit where $\nu \to 0$
and $\Delta_{\mathcal{O}} \to \frac{d}{2}$. The
computation should follow the same logic as presented in
the previous sections, although the radial wave-function
of the scalar field would involve $J_{0}(\Lambda x)$ and
the boundary expansion \eqref{bexp} has to be modified to
include log-terms, in which case the correct linear combinations
of Bessel functions satisfying the relevant boundary conditions
have to be computed separately. We will leave these interesting
possibilities for future work.

\section{Conclusion and discussion}
\label{sec:conclusion}

We have studied in the AdS space the 1-loop correction to the
boundary-boundary2-point function of a $U(1)$ gauge field coupled to a
complex scalar. The scalar is subjected to quadratic boundary
perturbations and satisfies \emph{mixed} boundary condition.
These boundary perturbations are related to double-trace
perturbations in the dual CFT.
While we concentrated our effort toward studying  the imaginary part of
the diagrams, which is of particular interest physically
and can be readily extracted via a simple generalization of Cutkosky
rules \cite{Chalmers:1998wu},
we found that in Poincare coordinates these loops suffer from
extra divergences shared by both the
real and imaginary parts of the diagram, arising
when the vertices of the diagrams are pushed toward the AdS \emph{horizon}
simultaneously and the geodesic distance between them shrinks to zero.
When the loop integral is reduced to a integral of both vertex
positions along the radial direction which gives rise to
a product of two double variable hypergeometric functions
(the Apell function $F_4$), the divergence manifests itself as singularities
along the boundary in the phase space where the momenta
of the photon and the loop scalar fields become collinear.

Fortunately, we manage to derive an explicit representation for these
divergences and discover that they are independent of the boundary couplings,
or in other words, the mixed boundary condition for the complex scalar.
While $F_4$ is known only as a power series,
and the singularities that concern us occur precisely along the boundary of
its convergent domain, we were able to extract
the exact properties of its singularities nevertheless
using simply the asymptotic expansion of the radial wave-functions
of the propagator.
It is found that the order of divergences are finite
and is related to the spacetime dimension in a simple algebraic way.
The coefficients are easily
computed as well. Most curiously, the leading divergence turns out to be
universal, depending only on the spacetime dimension, whereas the first
sub-leading divergence depends upon the spacetime dimension and
the bulk mass of the scalar field only.
The method presented can be applied to analyze the singularity structures
of other functions that have a similar integral form, such as similar
vertices in other curved spacetime background.

These results are readily applicable to certain physical problems.
In particular, we applied AdS/CFT correspondence to condensed matter
systems and studied the effects of random disorder on conductivity,
using the replica trick \cite{Fujita:2008rs}.
The effect of the disorder in this setup is captured by double-trace
boundary perturbations, corresponding precisely to mixed boundary condition
on the AdS side for the complex scalar. The dissipative part of the
current-current correlation in the CFT
is given by the imaginary part of the photon boundary-boundary
correlator in the AdS space, which begins to exhibit the effect of the
disorder from
the photon coupling to the particular charged scalar at 1-loop.
Since the singularities of the relevant loop integral are found to be
universal, we are able to subtract the results of the
conductivities obtained at various disorder coupling by the
corresponding conformal result, and obtain a finite difference. We
found that in general the disorder reduces the conductivity. Also,
since the disorder we consider is relevant in the infrared, we
find that generally in the extreme low frequency limit, the
loop correction begins to overwhelm the planar contribution,
suggesting the need for re-summation, but allows for the
possibility of the removal of the delta-function in the conductivity
known to exist at zero frequency due to momentum
conservation\cite{Hartnoll:2009sz}
even though the effect of the disorder appears at first sight to
be loop-suppressed.

Let us mention a couple of words on some other interesting
observations and conjectures.
We found that the same divergence in the limit as
loop vertices approach the horizon is
quite general in loop integrals in AdS space,
particularly when fields of higher spins,
such as photons and gravitons are involved, although they most likely
require more intriguing regularizations.
However, for loops involving only scalar $\phi^n$ vertices,
the absence of divergence is correlated to renormalizability
of the theory concerned.  It is also interesting that the
photon-fermion-fermion vertex displays a singularity near the horizon
with precisely the same power law behavior as the photon-scalar-scalar
vertex. It would be interesting to
understand the precise physics of these divergences.
In the Poincare patch where a mass gap is absent, we have, correspondingly
in the bulk a continuum spectrum of ``Kaluza-Klein" mass $\Lambda$ that
can be arbitrarily small, which might be intimately related
to the collinear divergence in the decay amplitude of the photon
as we have observed.

Yet, it is suggested in \cite{Skenderis:2008dg} that in real-time
computations one should build wave-functions
such that they vanish at the horizon. This would mean that
waves could only propagate as wave-packets rather
than as momentum eigenstates along isometry directions.
It is probable that the divergence can be cured by building
a wave packet that vanishes in the horizon limit but
asymptotes to a plane wave near the boundary.
However, since the regular part of the diagram in
the collinear limit is heavily suppressed by the phase-space volume,
we expect that our results should not be too sensitive to
the near horizon behavior of the wave-functions. We leave these
important and rigorous endeavor for future work.

\section*{Acknowledgement}
We thank Jaume Gomis, Louis Leblond, John McGreevy, Massimo Porrati,
 Amit Sever,
Aninda Sinha and Kostas Skenderis for inspiring discussions.
We are particularly grateful to Dileep Jatkar, Juan Maldacena, Rob Myers,
Dori Reichmann, Yogesh  Srivastava and Tadashi Takayanagi for lengthy
discussions and for their many suggestions and comments. We also thank
Daniel Freedman and Arkady Tseytlin for answering our email enquiries.
Research at Perimeter Institute is supported by the Government of Canada
through Industry Canada and by the Province of Ontario through the
Ministry of Research \& Innovation.

\begin{appendix}
\section{Contour integral of the scalar bulk-to-bulk propagator}
\label{app:contour}
In the Euclidean signature, the scalar bulk-to-bulk propagator is given by:
\begin{equation}
\label{eq:app_integral}
\int_0^{+\infty}
\frac{\Lambda\ud\Lambda}{\Lambda^2+k^2}\frac{[J_\nu(\Lambda x)
+\tilde f \Lambda^{2\nu} J_{-\nu}(\Lambda x)]
[J_\nu(\Lambda y)+\tilde f \Lambda^{2\nu} J_{-\nu}(\Lambda y)]}
{1+\tilde f^2\Lambda^{4\nu}+2\tilde f\Lambda^{2\nu}\cos\nu\pi}\,.
\end{equation}
We denoted $\tilde f=(2\Lambda)^{2\nu} \Gamma(1-\nu)/[f\Gamma(1+\nu)]$ for
brevity.

We can evaluate this propagator by replacing it by a contour integral.
Let's assume $x>y$ for the moment.  Notice that
\begin{equation}
\label{eq:two_K}
\begin{split}
&J_\nu(\Lambda x)+\tilde f \Lambda^{2\nu} J_{-\nu}(\Lambda x)=\\
	&\quad\frac{1}{\pi i}(e^{-\nu\pi i/2}+\tilde f
\Lambda^{2\nu} e^{\nu\pi i/2})
	K_\nu(\Lambda x e^{-\pi i/2})
	-\frac{1}{\pi i}(e^{\nu\pi i/2}+\tilde f\Lambda^{2\nu} e^{-\nu\pi i/2})
			K_\nu(\Lambda x e^{\pi i/2})
\end{split}
\end{equation}
and
\begin{equation}
1+\tilde f^2\Lambda^{4\nu}+2\tilde f\Lambda^{2\nu}\cos\nu\pi
=(\tilde f \Lambda^{2\nu}+e^{\nu\pi i})(\tilde f\Lambda^{2\nu}
		+e^{-\nu\pi i})\,,
\end{equation}
so the original integral is also given by
\begin{equation}
\int_0^{+\infty}\frac{\Lambda\ud\Lambda}{\pi i(\Lambda^2+k^2)}
\left[\frac{K_\nu(-i\Lambda x)}{\tilde f \Lambda^{2\nu} e^{-\nu\pi i/2}
	+e^{\nu\pi i/2}}
	-\frac{K_\nu(i\Lambda x)}{\tilde f\Lambda^{2\nu} e^{\nu\pi i/2}
		+e^{-\nu\pi i /2}}
\right]\cdot(\textrm{y terms})\,.
\end{equation}
We define
\begin{equation}
I_+=\int_0^{+\infty}\frac{\Lambda\ud\Lambda}{\pi i(\Lambda^2+k^2)}
\frac{K_\nu(-i\Lambda x)}{\tilde f \Lambda^{2\nu} e^{-\nu\pi i/2}
	+e^{\nu\pi i/2}}\cdot (\textrm{y terms}),
\end{equation}
and
\begin{equation}
I_-=-\int_0^{+\infty}\frac{\Lambda\ud\Lambda}{\pi i(\Lambda^2+k^2)}
\frac{K_\nu(i\Lambda x)}{\tilde f \Lambda^{2\nu} e^{\nu\pi i/2}
	+e^{-\nu\pi i/2}}\cdot (\textrm{y terms}).
\end{equation}
Let us also denote $C_+$ the closed contour in the complex plane that
consists of the entire real axis and the infinitely large
semicircle in the \emph{upper} half plane, and $C_-$ the opposite
contour that consists of the real axis and the infinitely large semicircle
in the \emph{lower} half plane.

It's readily verified that if one substitutes
\begin{equation}
\label{eq:upper_choice}
\Lambda\rightarrow \Lambda e^{\pi i}
\end{equation}
one finds
\begin{equation}
\label{eq:I+I-}
I_+\rightarrow -I_-\,.
\end{equation}
Similarly, upon substituting
$\Lambda$ by $\Lambda e^{-\pi i}$, one finds $I_-$ becomes $-I_+$.
The two phase choices just mentioned, however, are mutually exclusive and
therefore the full integrand \eqref{eq:two_K} does not enjoy
any symmetry property as $\Lambda\rightarrow -\Lambda$.

Let's evaluate the following contour integral, making use of
equation \eqref{eq:upper_choice} and \eqref{eq:I+I-}:
\begin{equation}
\begin{split}
\oint_{C_+}&\; \frac{z\ud z}{\pi i(z^2+k^2)}
\frac{K_\nu(-iz x)}{\tilde f z^{2\nu} e^{-\nu\pi i/2}
	+e^{\nu\pi i/2}}
\cdot (\textrm{y terms})\\
=&\int_{R^+}\; \frac{\Lambda\ud\Lambda}{\pi i(\Lambda^2+k^2)}
\frac{K_\nu(-i\Lambda x)}{\tilde f \Lambda^{2\nu} e^{-\nu\pi i/2}
	+e^{\nu\pi i/2}}
\cdot (\textrm{y terms})\\
&+\int_{R_-}\; \frac{\Lambda\ud\Lambda}{\pi i(\Lambda^2+k^2)}
\frac{K_\nu(-i\Lambda x)}{\tilde f \Lambda^{2\nu} e^{-\nu\pi i/2}
	+e^{\nu\pi i/2}}
\cdot (\textrm{y terms})\\
&+\int_{C_+-R}\; \frac{\Lambda\ud\Lambda}{\pi i(\Lambda^2+k^2)}
\frac{K_\nu(-i\Lambda x)}{\tilde f \Lambda^{2\nu} e^{-\nu\pi i/2}
	+e^{\nu\pi i/2}}
\cdot (\textrm{y terms})\\
=&I_+ + I_-,
\end{split}
\end{equation}
which gives rise precisely to the full integral \eqref{eq:app_integral}.
Here $R^+$, $R^-$ and $R$ denote the positive, negative, and the full
real axis respectively.  The choice of the contour $C_+$ is dictated
by equation \eqref{eq:upper_choice}
as the phase of $\Lambda$ rotates smoothly from $0$ to $\pi$.
This contour integral can be evaluated, on the other hand, by picking
up the residue at the pole $\Lambda=k e^{\pi i/2}$.
Similarly, we can evaluate the contour integral along $C_-$:
\begin{equation}
\begin{split}
-\oint_{C_-}&\; \frac{z\ud z}{\pi i(z^2+k^2)}
\frac{K_\nu(i z x)}{\tilde f z^{2\nu} e^{\nu\pi i/2}
	+e^{-\nu\pi i/2}}
\cdot (\textrm{y terms})\\
=&-\int_{R^+}\; \frac{\Lambda\ud\Lambda}{\pi i(\Lambda^2+k^2)}
\frac{K_\nu(i\Lambda x)}{\tilde f \Lambda^{2\nu} e^{\nu\pi i/2}
	+e^{-\nu\pi i/2}}
\cdot (\textrm{y terms})\\
&-\int_{R_-}\;\frac{\Lambda\ud\Lambda}{\pi i(\Lambda^2+k^2)}
\frac{K_\nu(i\Lambda x)}{\tilde f \Lambda^{2\nu} e^{\nu\pi i/2}
	+e^{-\nu\pi i/2}}
\cdot (\textrm{y terms})\\
&-\int_{C_+-R}\; \frac{\Lambda\ud\Lambda}{\pi i(\Lambda^2+k^2)}
\frac{K_\nu(i\Lambda x)}{\tilde f \Lambda^{2\nu} e^{\nu\pi i/2}
	+e^{\nu\pi i/2}}
\cdot (\textrm{y terms})\\
=&I_+ + I_-\,.
\end{split}
\end{equation}
This contour integral picks the residue at the pole $\Lambda=k e^{-\pi i/2}$
and leads to an identical result.

If $x<y$, both contour integrals given above diverge, in which case
we should just exchange the role of $x$ and $y$ and carry out the
above analysis similarly.

In summary, we find the propagator \eqref{eq:app_integral} equals
\begin{equation}
I_++I_-=\frac{2 K_\nu(kx)\left[I_\nu(ky)
+\tilde f k^{2\nu} I_{-\nu}(ky)\right]\Theta(x-y)}
{1+\tilde f k^{2\nu}}+ x\leftrightarrow y\,.
\end{equation}

\section{Some more details on the singularities of $I_{\nu, f_1, f_2}$}
\label{app:sec3_details}
We want to study the divergences of the integrals of the form
\cite{bailey}
\begin{equation}
\begin{split}
I\equiv&\int_0^{+\infty}
z^{\frac{d}{2}} K_{\frac{d}{2}-1}(iz)
J_\mu(\Lambda_1 z) J_\nu(\Lambda_2 z)\ud z\\
	=&
\frac{\Gamma\left[1+\frac{1}{2}(\mu+\nu)\right]\Gamma
	\left[\frac{1}{2}(d+\mu+\nu)\right]\Lambda_1^\mu\Lambda_2^\nu}
{(i)^{d/2+1+\mu+\nu}\Gamma(\mu+1)\Gamma(\nu+1)}\,
	F_4[1+\frac{1}{2}(\mu+\nu),\frac{1}{2}(d+\mu+\nu);\, \mu+1, \nu+1;\,
\Lambda_1^2, \Lambda_2^2]\,,
\end{split}
\end{equation}
when $1-|\Lambda_1|-|\Lambda_2|\rightarrow 0$.
Within the region $|\Lambda_1|+|\Lambda_2|<1$, the Appell
hypergeometric function $F_4$ has a series expansion given by
\begin{equation}
F_4[a,b;\, c, d;\,
\Lambda_1^2, \Lambda_2^2]= \sum_{m, n=0}^{\infty}\frac{(a)_m
	(b)_n}{m! n! (c)_m(d)_n}
	\Lambda_1^{2m} \Lambda_2^{2n}\,,
\end{equation}
where the notation $(\,\cdot\,)_m$ is defined by
\begin{equation}
(a)_m\equiv \frac{\Gamma\left(a+m\right)}{\Gamma(a)}\,.
\end{equation}
This expansion fails to converge whenever
$1-|\Lambda_1|-|\Lambda_2|\rightarrow 0$.
One can, however, by choosing the integral contour of $z$ carefully,
extend the definition of $I$ to the full $(\Lambda_1, \Lambda_2)$
plane which only becomes singular when
$1\pm\Lambda_1\pm\Lambda_2\rightarrow 0$.

To isolate these singularities of $I$, we examine the asymptotic behavior
of its integrand as $z\rightarrow\infty$ where the Bessel functions have
an asymptotic representation given by
\begin{gather}
\label{eq:asymptotic_K}
K_{\frac{d}{2}-1}(iz)\approx \sqrt{\frac{\pi}{2iz}}\,e^{-iz}\,
\sum_{n=0}^\infty\frac{(\frac{d}{2}-1, n)}{(2i z)^n} \\
\label{eq:asymptotic_J}
J_{\mu}(z)\approx\sqrt{\frac{1}{2\pi z}}\left[
e^{i(z-\frac{\mu\pi}{2}-\frac{\pi}{4})}\sum_{n=0}^\infty
\frac{i^n(\mu, n)}{(2z)^n}
+e^{-i(z-\frac{\mu\pi}{2}-\frac{\pi}{4})}\sum_{n=0}^\infty
\frac{(\mu, n)}{i^n(2z)^n}
\right]\,, \qquad (-\pi<\arg z<\pi)\,.
\end{gather}
Here we defined notation
\begin{equation}
(\nu, n)\equiv\frac{\Gamma(\frac{1}{2}+\nu+n)}
{n! \Gamma(\frac{1}{2}+\nu-n)}\,,
\end{equation}
and it is useful to note that $(\nu, 0)=1$ and $(\nu, n)=(-\nu, n)$.
These expansions are only valid in the
$z$-plane when its phase angle is between $-\pi$ and $\pi$. Using
these formulae, we can easily derive an asymptotic expansion for
the full integrand of $I$ in the limit $z\rightarrow +\infty$
and formally evaluate the integral $I$ by integrating the resultant
expansion term by term:
\begin{equation}
\label{eq:full_asymp_I}
\begin{split}
I\sim&\frac{1}{2^{\frac{d}{2}}\sqrt{\pi\Lambda_1\Lambda_2}}
\sum_{n=0}^\infty\sum_{k=0}^n\sum_{l=0}^{n-k}\int_0^{\infty}
\ud z\,
\frac{\left(\frac{d}{2}-1, n-k-l\right)(\mu, k)(\nu, l)}
{(2z)^{n+\frac{3-d}{2}}\Lambda_1^k\Lambda_2^l}\\
&\qquad\qquad\cdot
\left(e^{-i(\theta_\mu+\theta_\nu+\frac{3\pi}{4})}i^{2(k+l)-n}
	e^{-i[1-(\Lambda_1+\Lambda_2)]z}
+e^{-i(\theta_\mu-\theta_\nu+\frac{\pi}{4})} i^{2k-n}
e^{-i[1-(\Lambda_1-\Lambda_2)]z}\right.\\
&\qquad\qquad\qquad \left.
+e^{-i(\theta_\nu-\theta_\mu+\frac{\pi}{4})}i^{2l-n}
e^{-i[1-(\Lambda_2-\Lambda_1)]x}
+e^{i(\theta_\mu+\theta_\nu+\frac{\pi}{4})}
i^{-n}e^{-i[1+(\Lambda_1+\Lambda_2)]x} \right)\\
&\frac{1}{2^{\frac{3}{2}}\sqrt{\pi\Lambda_1\Lambda_2}}
\sum_{n=0}^\infty\sum_{k=0}^n\sum_{l=0}^{n-k}
\frac{\left(\frac{d}{2}-1, n-k-l\right)(\mu, k)(\nu, l)
\Gamma(\frac{d-1}{2}-n)}
{2^n\Lambda_1^k\Lambda_2^l}\\
&\cdot
\left(e^{-i(\theta_\mu+\theta_\nu+\frac{(d-2)\pi}{4})}(-)^{k+l}
		[1-(\Lambda_1+\Lambda_2)]^{n+\frac{1-d}{2}}
+ e^{-i(\theta_\mu-\theta_\nu+\frac{d\pi}{4})} (-)^k
[1-(\Lambda_1-\Lambda_2)]^{n+\frac{1-d}{2}}\right.\\
&\quad\quad\left.
+ e^{-i(\theta_\nu-\theta_\mu+\frac{d\pi}{4})}(-)^l
[1-(\Lambda_2-\Lambda_1)]^{n+\frac{1-d}{2}}
+e^{i(\theta_\mu+\theta_\nu+\frac{d-2}{4})\pi}
[1+(\Lambda_1+\Lambda_2)]^{n+\frac{1-d}{2}}\right)\,,
\end{split}
\end{equation}
where we have defined the phase angle
\begin{equation}
\theta_\mu\equiv\frac{\mu\pi}{2}\,.
\end{equation}
As explained in details in section~\ref{sec:divergence}, the
``$\sim$'' sign above must be understood as indicating that both sides
equal apart from unknown finite functions. Therefore,
terms that are not singular on the right hand side of the ``$\sim$'' sign
are meaningless and should be discarded.
In the domain we are interested in, $0<\Lambda_{1,2}< 1$, and the
poles of $I$ are given by
\begin{equation}
\label{eq:full_expan_I_1}
\begin{split}
I_{\nu, \infty}(i, \Lambda_1, \Lambda_2)
	&\sim\frac{e^{-2i\theta_\nu}\Gamma\left(\frac{d-1}{2}\right)}
{2i\sqrt{2\pi\Lambda_1\Lambda_2}}
\left\{\frac{1}{[1-(\Lambda_1+\Lambda_2)]^{\frac{d-1}{2}}} +
	\left[\frac{d-1}{4}-\frac{(\nu,1)}{d-3}\left(\frac{1}{\Lambda_1}
				+\frac{1}{\Lambda_2}\right)\right]
	\frac{1}{[1-(\Lambda_1+\Lambda_2)]^{\frac{d-3}{2}}}\right\}\\
	&\qquad+O\left([1-(\Lambda_1+\Lambda_2)]^{-(d-5)/2}\right)
	+\textrm{finite things}\,.
\end{split}
\end{equation}
More generally if $0\le \Lambda_{1,2}<+\infty$, $I$ may contain
other poles. Say, if $1\pm(\Lambda_1-\Lambda_2)\rightarrow 0$, we find
in the same way that
\begin{equation}
\label{eq:full_expan_I_2}
\begin{split}
I_{\nu, \infty}(i, \Lambda_1, \Lambda_2)
	&\sim
	-\frac{\Gamma\left(\frac{d-1}{2}\right)}
{2\sqrt{2\pi\Lambda_1\Lambda_2}}
\left\{\frac{1}{[1-(\Lambda_1-\Lambda_2)]^{\frac{d-1}{2}}}+
	\left[\frac{d-1}{4}-\frac{(\nu, 1)}{d-3}\left(\frac{1}{\Lambda_1}
				-\frac{1}{\Lambda_2}\right)\right]
	\frac{1}{[1-(\Lambda_1-\Lambda_2)]^{\frac{d-3}{2}}}\right\}\\
	&\qquad+O\left([1-(\Lambda_1+\Lambda_2)]^{-(d-5)/2}\right)
	+\Lambda_1\leftrightarrow\Lambda_2+\textrm{finite things}\,.
\end{split}
\end{equation}
Since the asymptotic expansion \eqref{eq:asymptotic_K} and
\eqref{eq:asymptotic_J}are not valid along the negative axis of $z$,
this method can not be used to analyze the pole structures
if either $\Lambda_{1,2}$ becomes negative. In particular, the pole
of the form of $1/(1+\Lambda_1+\Lambda_2)^\alpha$ can not
be properly derived as above, but they can be inferred easily
by the symmetry properties of $I$
when $\Lambda\rightarrow -\Lambda$.

If, instead of the standard Bessel functions $J_{\pm\nu}$,
we wish to use $J_{\nu, f}$ as defined by \eqref{besself}
in the integral formulae for $I$ as we should
if the bulk scalar $\phi$ satisfies a mixed boundary condition,
we must replace the asymptotic expansion of $J_{\pm\nu}$ by
that of $J_{\nu, f}$. It's easily verified that
\begin{equation}
J_{\nu, f}(z)\overset{{z\rightarrow\infty}}{\approx}
\sqrt{\frac{1}{2\pi z}}\left[
e^{i(z-\theta_{\nu, f, \Lambda}-\frac{\pi}{4})}\sum_{n=0}^\infty
\frac{i^n(\mu, n)}{(2z)^n}
+e^{-i(z-\theta_{\nu, f, \Lambda}-\frac{\pi}{4})}\sum_{n=0}^\infty
\frac{(\mu, n)}{i^n(2z)^n}
\right]
\end{equation}
which differs from those of $J_{\pm\nu}$ by merely a less trivial
phase angle that we have denoted as $\theta_{\nu, f, \Lambda}$ above
and is defined implicitly through the following equation:
\begin{equation}
\tan \theta_{\nu, f,\Lambda}\equiv\tan \frac{1-\frac{\Lambda^{2\nu}}{f}
	\frac{\Gamma(1-\nu)}{\Gamma(1+\nu)}}
{1+\frac{\Lambda^{2\nu}}{f}
	\frac{\Gamma(1-\nu)}{\Gamma(1+\nu)}}\tan\theta_\nu\,.
\end{equation}
Obviously $\theta_{\nu, f}$ interpolates between $\theta_{\nu}$
and $\theta_{-\nu}=-\theta_{\nu}$ as $f$ varies from $\infty$ to $0$.

Therefore, it is straightforward to generalize the results given
above and find, for example, when $1-(\Lambda_1+\Lambda_2)\rightarrow 0$:
\begin{equation}
\begin{split}
I_{\nu, f_1, f_2}(i, \Lambda_1, \Lambda_2)
	&\sim\frac{e^{-i(\theta_{\nu, f_1, \Lambda_1}
			+\theta_{\nu, f_2, \Lambda_2})}
			\Gamma\left(\frac{d-1}{2}\right)}
{2i\sqrt{2\pi\Lambda_1\Lambda_2}}
\left\{\frac{1}{[1-(\Lambda_1+\Lambda_2)]^{\frac{d-1}{2}}}\right.\\
	&\left.\qquad+
	\left[\frac{d-1}{4}-\frac{(\nu,1)}{d-3}\left(\frac{1}{\Lambda_1}
				+\frac{1}{\Lambda_2}\right)\right]
	\frac{1}{[1-(\Lambda_1+\Lambda_2)]^{\frac{d-3}{2}}}\right\}\\
	&\qquad+O\left([1-(\Lambda_1+\Lambda_2)]^{-(d-5)/2}\right)
	+\textrm{finite things}\,.
\end{split}
\end{equation}
which is essentially identical to that of $I_{\nu,\infty}$
except that $\theta_{\nu, f,\Lambda}$ depends both on $f$ and $\Lambda$
now \footnote{In this case one can again set $p=1$ which amounts to
a scaling of $f$. Rigorously speaking, $\theta_{\nu, f,\Lambda}$
depends on both $f$ and $p$, but only through their ratio $f/p$.}.
This is, of course, necessary if any nontrivial results
are to be expected.

\end{appendix}

\end{document}